\documentclass{aa}
\usepackage{hyperref}
\hypersetup{colorlinks, allcolors=blue}
\usepackage{graphicx}    
\usepackage{color}  
\usepackage{txfonts}
\usepackage{inputenc}
\usepackage{multicol}
\usepackage{multirow}
\usepackage[table,xcdraw]{xcolor}
\usepackage{siunitx}
\DeclareSIUnit{\solarmass}{\text{M}_{\odot}}
\DeclareSIUnit{\parsec}{pc}
\DeclareSIUnit{\year}{yr}
\DeclareSIUnit{\mag}{mag}
\DeclareSIUnit{\arcsec}{arcsec}
\usepackage{amsmath}
\usepackage{xcolor}
\usepackage{booktabs}
\usepackage{color} %DIF > 

\usepackage{algorithm} 
\usepackage{algpseudocode} 
\begin{document}

   \title{
   Toward the fabric of the Milky Way
   }
        \subtitle{I. The density of disk streams from a local 250$^3$ pc$^3$ volume\thanks{The full source catalog described in Tables 1 and F.1 is only available at the CDS via anonymous ftp to \href{http://cdsarc.u-strasbg.fr}{cdsarc.u-strasbg.fr}  (\href{ftp://130.79.128.5/}{130.79.128.5}). Figure E.5 and additional figures are available online via Zenodo: \href{https://doi.org/10.5281/zenodo.14278685}{https://doi.org/10.5281/zenodo.14278685}.}}

   \author{Sebastian Ratzenb\"ock\inst{1,2,3}
          \and
          Jo\~ao Alves\inst{1,2}
          \and
          Emily L. Hunt\inst{4}
          \and
          N\'uria Miret-Roig\inst{1}
          \and
          Stefan Meingast\inst{1}
          \and
          Torsten M\"oller\inst{2,3}
          }

   \institute{University of Vienna, Department of Astrophysics,
              T\"urkenschanzstra{\ss}e 17, 1180 Vienna, Austria\\
              \email{sebastian.ratzenboeck@univie.ac.at}
         \and
             University of Vienna, Research Network Data Science at Uni Vienna, Kolingasse 14-16, 1090 Vienna, Austria
        \and 
            University of Vienna, Faculty of Computer Science, W\"ahringer Straße 29/S6, 1090 Vienna, Austria
        \and
            Landessternwarte, Zentrum f\"ur Astronomie der Universität Heidelberg, K\"onigstuhl 12, 69117 Heidelberg, Germany
    }
   \date{Submitted to A\&A on Aug. 12, 2024}
 
  \abstract
  {We studied 12 disk streams found in a 250$^3$ pc$^3$ volume in the solar neighborhood, which we define as coeval and comoving stellar structures with aspect ratios greater than 3:1. Using \emph{Gaia} Data Release 3 data and the advanced clustering algorithms \texttt{SigMA} and \texttt{Uncover}, we identified and characterized these streams beyond the search volume, doubling, on average, their known populations. We estimate the number density of disk streams to be $\approx 820$ objects\,/\,kpc$^3$ (for $|Z| < 100$\,pc), or surface densities of $\approx 160$ objects\,/\,kpc$^2$. These estimates surpass N-body estimates by one to two orders of magnitude and challenge the prevailing understanding of their destruction mechanisms. Our analysis reveals that these 12 disk streams are dynamically cold with 3D velocity dispersions between 2 and 5\,km\,s$^{-1}$, exhibit narrow sequences in the Hertzsprung-Russell diagram, and are highly elongated with average aspect ratios of 7:1, extending up to several hundred parsecs.  We find evidence suggesting that one of the disk streams, currently embedded in the Scorpius-Centaurus  association, is experiencing disruption, likely due to the primordial gas mass of the association. 
}


   \maketitle

\defcitealias{Ratzenboeck:2023a}{Paper\,I}
\defcitealias{Ratzenboeck2020}{RMA20}
\defcitealias{Hunt:2023}{HR23}
\defcitealias{Kounkel2019}{KC19}
\defcitealias{Qin:2023}{QZTC23}
\defcitealias{Kamdar:2021_stream}{HCT21}
\defcitealias{Meingast2019}{MAF19}
\defcitealias{Meingast2021}{MAR21}

%
%__________________________________________________________

\section{Introduction}

Disk streams are coeval and comoving elongated stellar structures within the Milky Way's disk. They are the disk counterparts of halo streams, which are traditionally studied in the context of the Milky Way halo \citep[e.g.,][]{Grillmair:1995, Odenkirchen:2001, Malhan:2018, Bovy:2016, Ibata:2016, PriceWhelan:2018}. Unlike halo streams, which are primarily associated with tidally disrupted globular clusters and dwarf galaxies accreted by the Milky Way, disk streams are believed to originate from the disruption of bound and unbound clusters (associations) born in the Milky Way \citep[e.g.,][]{Eggen1996-da,Meingast2019,Kamdar:2019_sim, Meingast2021,Kamdar:2021_stream}. Disk streams, in particular nearby ones, constitute excellent laboratories for studying planet formation \citep[e.g.,][]{Curtis2019-es,Newton2021-zt} and can provide insights into processes such as the dissolution of star clusters, the influence of Galactic dynamics, and the interaction between stellar structures and giant molecular clouds (GMCs).

The remarkable precision of ESA's \emph{Gaia} data within the local Milky Way has revolutionized our ability to uncover disk streams, an endeavor that previously seemed near impossible due to their subtle presence against the densely populated backdrop of the Milky Way disk. The first bona fide disk stream, Meingast-1 (Pisces-Eridanus; \citealt{Meingast2019}), is a coeval and comoving unbound structure with an age of about 120~Myr \citep{Curtis2019-es}, a length of at least 400 pc, and a vertical extent of only about 50 pc, covering about $120^\circ$ of the sky. This stellar structure was the first chemically homogeneous disk stream identified on the Milky Way disk \citep{Ratzenboeck2020, Hawkins:2020}.

While there is currently no working definition of a disk stream, in this work we define it as a coeval and comoving stellar structure with aspect ratios (between the first and third principal components in \emph{XYZ}) greater than 3:1. This broad definition includes cluster tidal tails \citep[e.g.,][]{Roeser2019,MeingastAlves2019,Jerabkova:2021,Kroupa2022-to}, unbound clusters (associations), young moving groups, young local associations \citep[e.g.,][]{GagneFaherty2018,Beccari2020,Miret-Roig2020,Tian2020-bv}, and cluster coronas \citep{Meingast2021,Moranta:2022}. This working definition is appropriate at this stage, as it keeps us from overclassifying the outcome of different complex processes (e.g., Galactic tidal forces, differential rotation, and initial star-forming gas configurations). We leave a discussion on the relative role of the different formation processes, and a better disk stream definition, to be tackled whenever a statistically significant sample of disk streams becomes available.

Notwithstanding recent advancements in the detection of initial disk streams, their potential utility as observational laboratories for the processes governing planet formation and the genesis of the Galactic field population remains largely unexamined. This is due to the lack of a census of disk streams in the Milky Way. While we know of about 100 halo streams \citep{Mateu2023-rf}, only a handful of disk streams have been identified. This limits our understanding of the processes transforming clustered young stellar populations into the Galaxy's field population. To address this, we need advanced tools to identify coeval and comoving populations down to densities well below the average volume density of stars in the Milky Way.

In this work we focus on establishing the volume and surface density of disk streams in the local Milky Way. We studied 12 disk streams within a fully sampled 250$^3$ pc$^3$ local volume. These 12 disk streams were identified as interloper structures in the Scorpius-Centaurus (Sco-Cen) study \citet[hereafter \citetalias{Ratzenboeck:2023a}]{Ratzenboeck:2023a}, which employed the \texttt{SigMA} algorithm. Here, we search for additional members of these disk streams outside the initially defined search box of 250$^3$ pc$^3$, using the \texttt{Uncover} algorithm from \citet{Ratzenboeck:2023c} to characterize their basic physical properties, such as mass and age. Our volume-complete sample will allow us to provide a first estimate of the abundance and lifetime of disk streams in the local Milky Way disk. 

%__________________________________________________________________

\section{Data}\label{sect:data}
This study uses position and velocity information from the \textit{Gaia} Data Release 3 \citep[DR3;][]{Vallenari2022}. We selected all sources within 500\,pc that pass the following quality criteria, as defined in \citetalias{Ratzenboeck:2023a}:
\begin{equation}\label{eq:data_selection}
    \begin{split}
        \varpi / \sigma_{\varpi}  &> 4.5. \\
        \varpi &> 0.
    \end{split}
\end{equation}

We determined the distance by inverting the parallax measurement. Due to the large parallax signal-to-noise criterion and relatively small distances (<500\,pc), inverted parallaxes are in good agreement with distance predictions from, for example, \citet{Bailer-Jones2021} but also avoid introducing a space distribution prior that may smooth out real substructure that we wish to find. Combined with \emph{Gaia} DR3 right ascension (ra, deg) and declination (dec, deg), we determined 3D space positions in the heliocentric Galactic Cartesian coordinate frame \emph{XYZ} (in pc). After applying the quality criteria in Eq. (\ref{eq:data_selection}), our final dataset contains 25\,475\,384 sources, referred to as the 500\,pc search data. We also selected a subsample of the 500\,pc search data (hereafter, 6D search data) with an absolute radial velocity error of less than two: $\sigma_{v_r} < 2$\,km\,s$^{-1}$. This sample contains 2\,078\,715 sources within 500\,pc for which we computed precise 3D space motions in heliocentric Galactic Cartesian coordinates \emph{UVW} (in km\,s$^{-1}$). We find no systematic differences in the resulting streams' morphology (length, location, velocity dispersion) when clustering in the Galactocentric cylindrical velocity coordinates ($v_{R}, v_{\phi}, v_{z}$) instead and only percent level differences in the identified stream sizes. Since the impact of these two reference frames on our pipeline is negligible, we opted to use the heliocentric Galactic Cartesian coordinate system.

This study focuses on 12 disk streams identified as interlopers by the \texttt{SigMA} clustering pipeline in \citetalias{Ratzenboeck:2023a}. \citetalias{Ratzenboeck:2023a} aimed to investigate the sub-structuring of the Sco-Cen association, identifying distinct Sco-Cen subpopulations as well as 48 additional stellar clusters that were not kinematically related to Sco-Cen and thus excluded from further discussion. Among these 48 unrelated clusters, this study follows up on 12 populations that exhibit significant prolate morphologies with pronounced elongations, meeting our aspect ratio criterion of >3:1. The remaining 36 clusters, classified as established open clusters or moving groups, do not display the morphological or kinematic characteristics consistent with disk streams and are therefore excluded from this analysis.

These 12 disk streams are located within the search box defined in \citetalias{Ratzenboeck:2023a} with dimensions \emph{X}: $[-50, 250]$, \emph{Y}: $[-200, 50]$, and \emph{Z}: $[-95, 100]$\,pc. This corresponds to a box size of 300\,pc $\times$ 250\,pc $\times$ 195\,pc, encompassing a total volume of 14\,625\,000\,pc$^3$. For simplicity and clarity, we approximate and refer in this manuscript to this volume as a cube with 250\,pc sides (i.e., a 250$^3$ pc$^3$ box\footnote{A more accurate equivalent is a cube with sides of $\sim$ 244\,pc. For ease of notation, we have rounded this value to the nearest 50\,pc. Naturally, all volume and surface density calculations are done using the true box volume of 14\,625\,000\,pc$^3$.}). Most of the 12 streams appear to terminate abruptly at the edges of this box, suggesting its boundaries truncate them. To recover the potential missing members of these truncated clusters, we expanded the search to the full 500\,pc dataset, utilizing a combination of precise phase-space information (\emph{XYZ} + \emph{UVW}) and its projected 5D counterpart using \emph{XYZ} and tangential velocities $v_{\alpha}$ and $v_{\delta}$ (in km\,s$^{-1}$) for cases where radial velocity data are unavailable.

\section{Method}\label{sect:method}
\begin{figure*}
    \centering
    \includegraphics[width=0.97\textwidth]{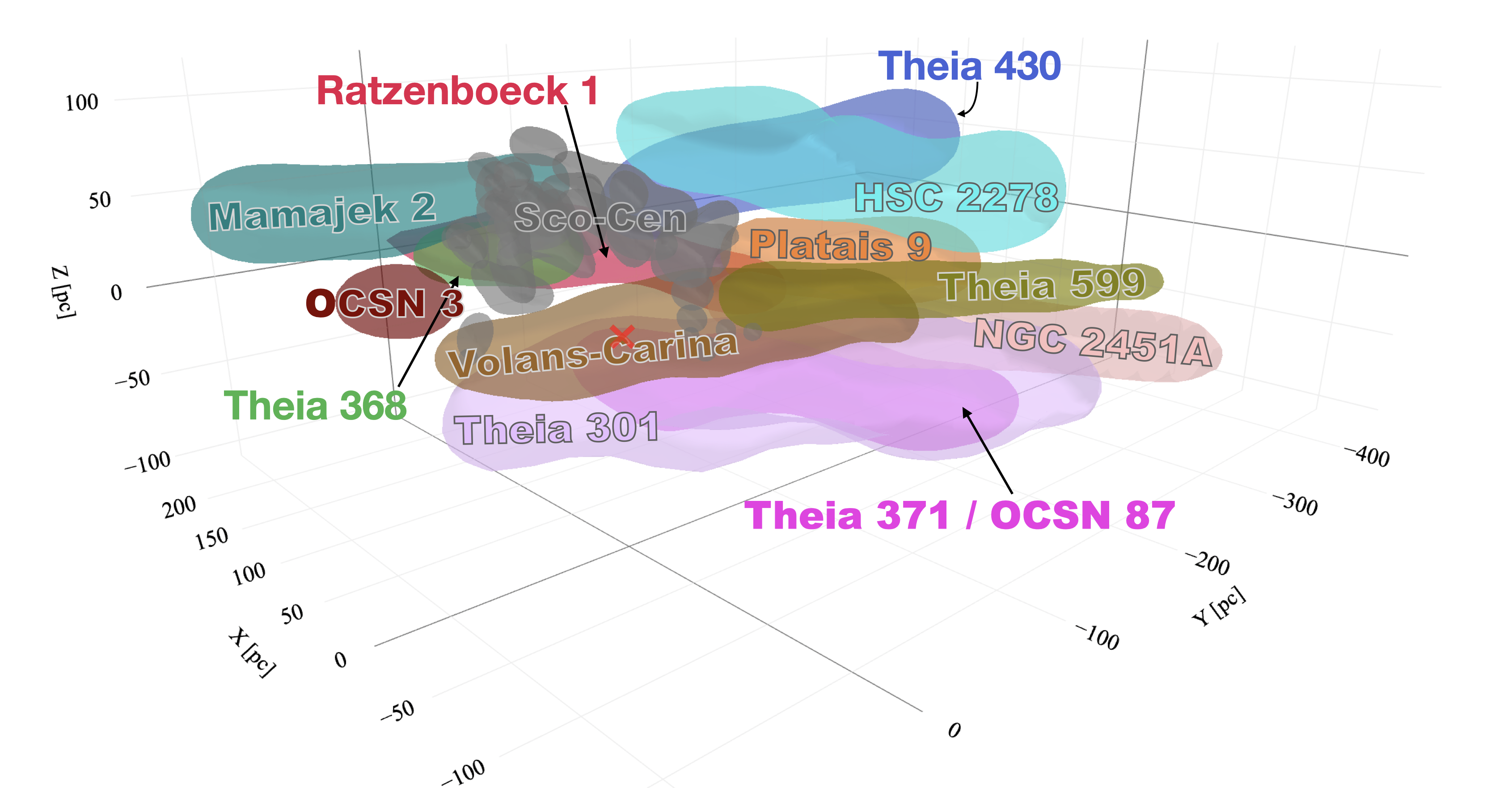}
    \caption{3D distribution of 12 disk streams (in color) alongside Sco-Cen from\,\citet{Ratzenboeck:2023b}. The Sun is at (0,0,0) and is represented by the red ``x.''
    % and Z = 0  ...
    For better visualization, see the link to the \href{https://homepage.univie.ac.at/sebastian.ratzenboeck/wp-content/uploads/2024/07/disk_streams_ratzenboeck2024.html}{interactive 3D version} of this figure, which allows the user to toggle on and off individual sources and the initial search box of 250$^3$\,pc$^3$.}
    \label{fig:3d_view}
\end{figure*}

We used a combination of two automated search tools, \texttt{SigMA} and \texttt{Uncover}\,\citep[]{Ratzenboeck2020}, for the stream membership analysis. In this study, we employed \texttt{SigMA} to recover the full extent of each progenitor stream using 6D phase space data. Given a 6D selection, \texttt{Uncover} was used to identify members without (precise) radial velocity measurements. 

Due to its size, our search strategies (see Sects.\,\ref{sect:sigma} and \ref{sect:uncover}) cannot directly apply to the 500\,pc search data. Consequently, we adopted a targeted approach that individually examines the 12 progenitor streams within a more manageable local selection of the abovementioned datasets. We discarded sources in the position and velocity space that are likely unrelated with progenitor stream. To do so, we first computed the three principal axes of the progenitor stream by applying principal component analysis to their \emph{XYZ} coordinates. Then, we determine the extent $\Delta x_i$ of each progenitor stream along its principal axes by identifying the minimum and maximum positions of sources projected onto each axis. The box within which we search for additional stream members is chosen to extend $3 \times \Delta x_i$ out from the progenitor stream’s median \emph{XYZ} position along each principal axis, with $i \in \{1, 2, 3\}$. Thus, we limit the streams’ sizes to expand by a factor of 6. The extent of additional members we recover across all 12 streams is well constrained within their respective search boxes and does not come close to any border.

To improve the contrast of the populations against the background field in \emph{XYZ}, we remove unlikely members through a kinematic selection. To this end, we compute the median Galactic Cartesian 3D velocity $v$ for each stream and retain sources in (local versions of) the 6D search data if a source's velocity difference to $v$ is less than 20 km\,s$^{-1}$. Sources beyond this velocity cut, concerning, for example, members with large radial velocity uncertainties, can still enter the final selection through our two-step selection pipeline detailed in Sects. \ref{sect:sigma} and \ref{sect:uncover}. This velocity selection is still useful, though, as the blind search (see Sect.~\ref{sect:sigma}) becomes empirically more sensitive to the low-contrast tails of the streams.

\subsection{Expanding the progenitor stream’s extent}\label{sect:sigma}
We search for overdensities in 6D phase space using the \texttt{SigMA} pipeline to identify each stream in its respective local 6D search data. \texttt{SigMA} is an unsupervised hierarchical density-based learning method that identifies clusters as statistical overdensities in input space (here, 6D phase space) separated by regions of significantly lower density. For a detailed description of the \texttt{SigMA} pipeline, we refer to \citetalias{Ratzenboeck:2023a}.

Due to their size and relative proximity, the 12 progenitor streams cover huge areas on the sky, with angular extents of about $90^{\circ}$ on average and as large as $170^{\circ}$. Even with distinguished compact overdensities in \emph{UVW}, projection effects can cause highly dispersed and non-convex distributions in tangential velocity space, significantly hindering detection capabilities with density-based clustering tools. Thus, in contrast to previous applications of \texttt{SigMA}, we ran the pipeline directly on the 6D search data to tailor the pipeline to these large structures.
We adopt the parameter choices discussed in \citetalias{Ratzenboeck:2023a}, except for the appropriate scaling factor values, which are affected by the change of the input space from the 5D to the 6D phase space. We update the kinematic scale factors (see Sect. 3.3.3 in \citetalias{Ratzenboeck:2023a}) using the selection of progenitor stream members (see Appendix\,\ref{app:sigma_parameters} for a detailed description). 

Due to the large box sizes, running \texttt{SigMA} in each of the 12 local 6D datasets results in multiple recovered overdensities. The population of interest is determined as the cluster maximizing the crossmatch rate with the progenitor stream. Except in one case, which exhibits a possible overlap with the Coma-Ber cluster, we find a clean and unique match between resulting \texttt{SigMA} groups and progenitor streams. We discuss this overlap in more detail in Sect.\,\ref{sect:results}.

\subsection{Identifying unknown members}\label{sect:uncover}
The resulting 6D \texttt{SigMA} selection discussed in Sect.\,\ref{sect:sigma} is subsequently fed into the \texttt{Uncover} pipeline to identify unknown stream members for which no (precise) radial velocities are available. \texttt{Uncover} is an extended membership analysis technique that integrates known members of stellar populations to identify undetected members. 

\texttt{Uncover} merges a powerful black-box algorithm, one-class support vector machines \citep[OCSVM;][]{Schoelkopf:1999}, into a statistical framework to provide meaningful parameter selection tools and improve membership accuracy. For a detailed description of the \texttt{Uncover} pipeline, we refer to \citet{Ratzenboeck2020}, \citet{Grasser2021}, and \citet{Ratzenboeck:2023c}. Instead of manually choosing OCSVM model parameters, \texttt{Uncover} employs ranges of interpretable summary statistics the trained model must adhere to, such as estimates on the number or the maximally allowed velocity dispersion of yet unseen members. We adopted the parameter selection approach presented in \citet{Ratzenboeck2020} due to the similarity in the presented use case to ours and refer the reader to this manuscript for an in-depth discussion on parameter choices.

We applied \texttt{Uncover} to the local 5D search box (using the features \{\emph{X}, \emph{Y}, \emph{Z}, $v_{\alpha}, v_{\delta}$\}) where the candidate members identified in the previous clustering step were used as the training set. The final inferred member selection remains remarkably stream-like, with all 12 populations appearing highly elongated along their bulk velocity direction. Inside the original search box, encompassing a volume of $250^3$\,pc$^3$, these 12 disk streams are tightly packed and appear to envelop Sco-Cen. 

\begin{figure*}
    \centering\includegraphics[width=0.95\textwidth]{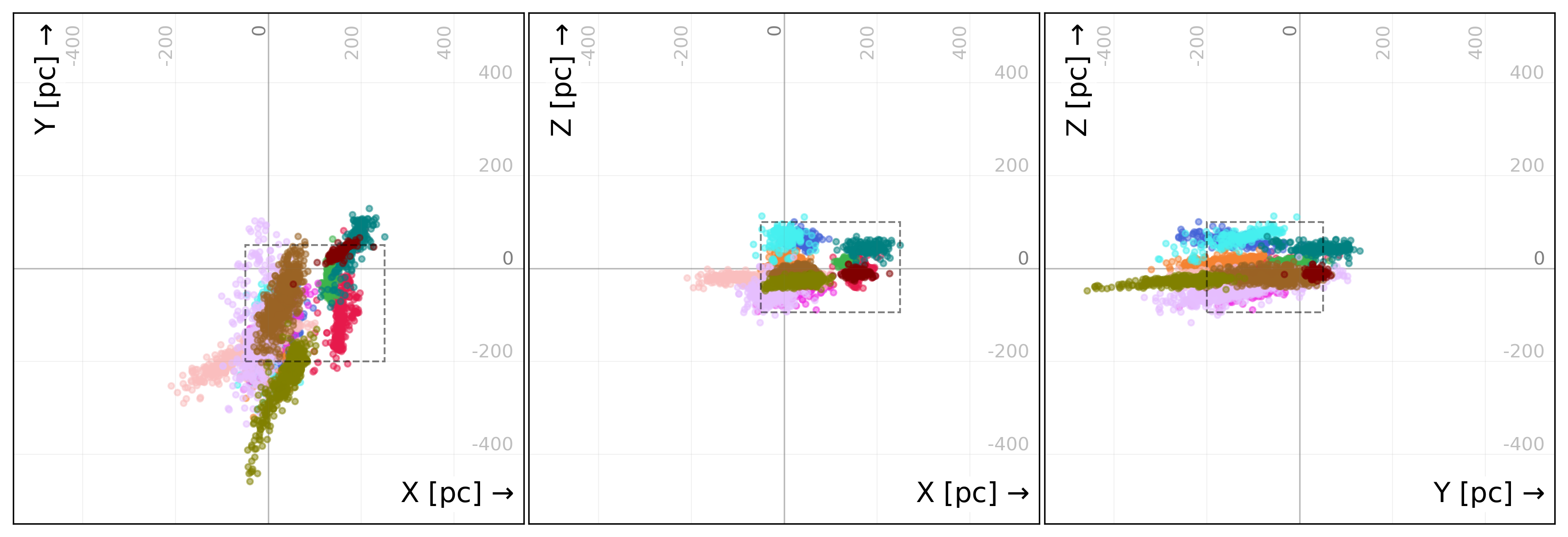}
    \caption{Spatial distribution of our selection for 12 disk streams in heliocentric Galactic coordinates. Colors have the same meaning as in Fig.~\ref{fig:3d_view}. Because of sensitivity, the elongations of these disk streams are lower limits to their true elongation. The dashed rectangle indicates the search volume of approximately 250$^3$\,pc$^3$ within which we aim to identify the local disk stream population. Many of the progenitor streams identified extend far beyond the initial search box.}
    \label{fig:xyz}
\end{figure*}

\section{Results}\label{sect:results}
Table~\ref{tab:group_stats} summarizes the 12 recovered disk streams. We assigned each object to an increasing stream identifier (SID) from S1 to S12 and report the name assigned by crossmatching our sources to the literature. 
Except for disk stream S1, we find a sufficiently good match between the two population morphologies in the literature (see Sect.~\ref{sect:comparison}). Since object S1 is unknown, we name it ``Ratzenboeck~1.''  In the following, we discuss the properties of the identified streams in more detail. 

\subsection{Spatial distribution}
Figure~\ref{fig:3d_view} shows the positional extent of the 12 identified disk streams in Galactic Cartesian coordinates (\emph{XYZ}). Disk streams are displayed via simplified colored shapes along with Sco-Cen's subcluster population, represented by gray shapes \citep[see][]{Ratzenboeck:2023b}. Figure~\ref{fig:3d_view} highlights the scale of the recovered disk streams whose size exceeds that of Sco-Cen by up to a factor of 3. Although the entire Sco-Cen population extends across 150\,pc in the \emph{X-Y} plane, the length of the identified disk streams ranges between $120$ and $430$\,pc with an average length of $280$\,pc. Furthermore, all clusters are highly prolate, meaning the median absolute deviation (MAD) along one principal axis is significantly larger than the MAD along the other two principal axes. The MAD is used here to provide a reliable and robust measure of the spread along each principal axis. All 12 disk streams have aspect ratio measurements (ratio between largest to smallest principal component axis) ranging from $3.3$ to $10.1$ with an average aspect ratio of $7.1$. Additionally, Table~\ref{tab:group_stats} provides MAD ratios among the three principal axes in \emph{XYZ} normalized to the smallest component, showcasing the prolate nature of the identified populations. However, we note that the provided aspect ratio values likely constitute lower limits, and we expect these structures to grow in length with improved precision of upcoming \emph{Gaia} data releases. 

Figure~\ref{fig:xyz} displays the individual recovered sources color-coded by disk stream membership. The boxes are chosen to retain an equal aspect ratio, highlighting the streams' extents of up to a few hundred parsecs in the \emph{X-Y} plane compared to their compact size along the \emph{Z} dimension. In the \emph{X-Y} plane, the disk streams cover different inclination angles between the \emph{X} and \emph{Y} axis ranging from $45^{\circ}$ to $90^{\circ}$. 

\subsection{Velocity distribution}\label{sect:result_vel}
The populations are dynamically cold, with 3D velocity dispersions between $2.1$ and $5.1$\,km\,s$^{-1}$. These values were determined using a robust deconvolution method that simultaneously addresses large radial velocity measurement errors and accounts for outliers (see Appendix~\ref{app:vel_disp} for details). This approach reduces the estimated 3D velocity dispersion by a factor of \textasciitilde5\ when compared to traditional empirical covariance estimates. Figures \ref{fig:uvw_1} -- \ref{fig:uvw_3} illustrate the distribution of sources in Cartesian velocity space (\emph{UVW}). To mitigate the risk of underestimating 3D velocity dispersions, we also calculated them using the MAD. While this robust measure effectively accounts for potential outliers, it does not (inherently) correct for large radial velocity measurement errors. Consequently, it yields slightly higher velocity dispersion estimates, ranging from 2.7 -- 7.9\,km\,s$^{-1}$. Both estimates are presented in Table~\ref{app:vel_disp} where the MAD estimate is indicated in parentheses. (For a discussion on the velocity dispersion estimation, see Sect.~\ref{sect:discussion_vel_disp} and Appendix~\ref{app:vel_disp}.)

\subsection{Result validation}
Since the recovered disk streams cover large volumes with many co-spatial stars, the chance of random field contaminants is much higher than, for example, in more compact open cluster configurations. To validate our clustering pipeline, we applied three different contamination estimation procedures. 

First, our deconvolution approach enables us to quantify the contamination rate in each sample by incorporating a dedicated outlier component into a two-component Gaussian mixture model (GMM). This component accounts for both ``true'' outliers and artificial kinematic outliers caused by binaries or large radial velocity uncertainties. While its primary purpose is to prevent the artificial inflation of the signal component, the inferred mixture weight of the outlier component can also be directly interpreted as the fraction of random outliers in the sample. Using this procedure, we find an average contamination rate across the 12 disk streams of $9\%$, with individual values ranging between $5$ and  $17\%$ (see Appendix~\ref{app:vel_disp}). 

Second, we compared a background population of co-spatial sources that share the same volume (in \emph{XYZ}) with our selected stream members in the observational Hertzsprung-Russell diagram (HRD). We find that the distribution of identified stream members represents a significantly narrower configuration around selected model isochrones than a sample from the background population. This provides substantial evidence that the disk stream members we identified constitute coeval populations (see Appendix~\ref{app:contamination}).

Third, we compared the velocity distribution of recovered disk streams to that of the corresponding background sources. By determining the number density of stream members and contrasting it with the expected number density of background sources moving with the stream’s bulk motion, we derived a signal-to-noise ratio (S/N) and contamination estimate for each stream. We derive a mean S/N of 28, with values ranging from $5$ to $114$, and estimate a contamination level of $7 \pm 4\%$. We provide the estimated S/N in Table~\ref{tab:group_stats} (see Appendix~\ref{app:contamination}).

Although kinematically distinct, several disk streams exhibit partial spatial overlap with their respective members, which are intermixed to varying degrees. Notably, we find three groups: (1) Theia 430 and HSC 2278, (2) Theia 371/OCSN 87, NGC 2451A, Theia 301, and OCSN 3, and (3) a group made up of disk streams Ratzenboeck~1, Theia 368, and Mamajek 2. Members of the third group are found to also spatially coexist, at varying degrees, with members of the Sco-Cen association, as discussed in Sect.~\ref{sect:scocen}. These spatial overlaps raise the possibility of shared origins among these structures. However, a detailed kinematic analysis reveals that in all but one case, these co-spatial streams show statistically significant differences in their velocities and ages. This analysis is further detailed in Appendix~\ref{app:kinematic_pop_analysis} and Sect.~\ref{sect:discussion_vel_disp}, where we discuss a case that warrants further scrutiny, Ratzenboeck~1 and Theia 386.

Compared to previous unsupervised studies (not including targeted searches such as \citealt{Meingast2021}), our pipeline identifies, on average, more than twice as many candidate members (see Sect.~\ref{sect:comparison}). Our pipeline allows us to detect median stream densities (number of sources divided by the entire population volume) of 1 star per $10^3$\,pc$^3$ (or 0.001 stars/pc$^3$), which is about 50 times lower than the surrounding field density. At both extremes, average stream densities are an order of magnitude apart. Whereas the densest structures, such as NGC 2451A and Platais 9, have average densities (across the entire population) of about 2 -- 3 sources per $10^3$\,pc$^3$, we reach average densities as low as 0.2 -- 0.5 sources per $10^3$\,pc$^3$ (min 0.0002 stars/pc$^3$) across the entire disk steam in HSC 2278, Theia 371/OCSN 87, Theia 301, Theia 430, and Ratzenboeck~1, effectively resolving structures 250 times below the field density in \emph{XYZ} coordinates.

\begin{figure*}
    \centering
    \includegraphics[width=0.95\textwidth]{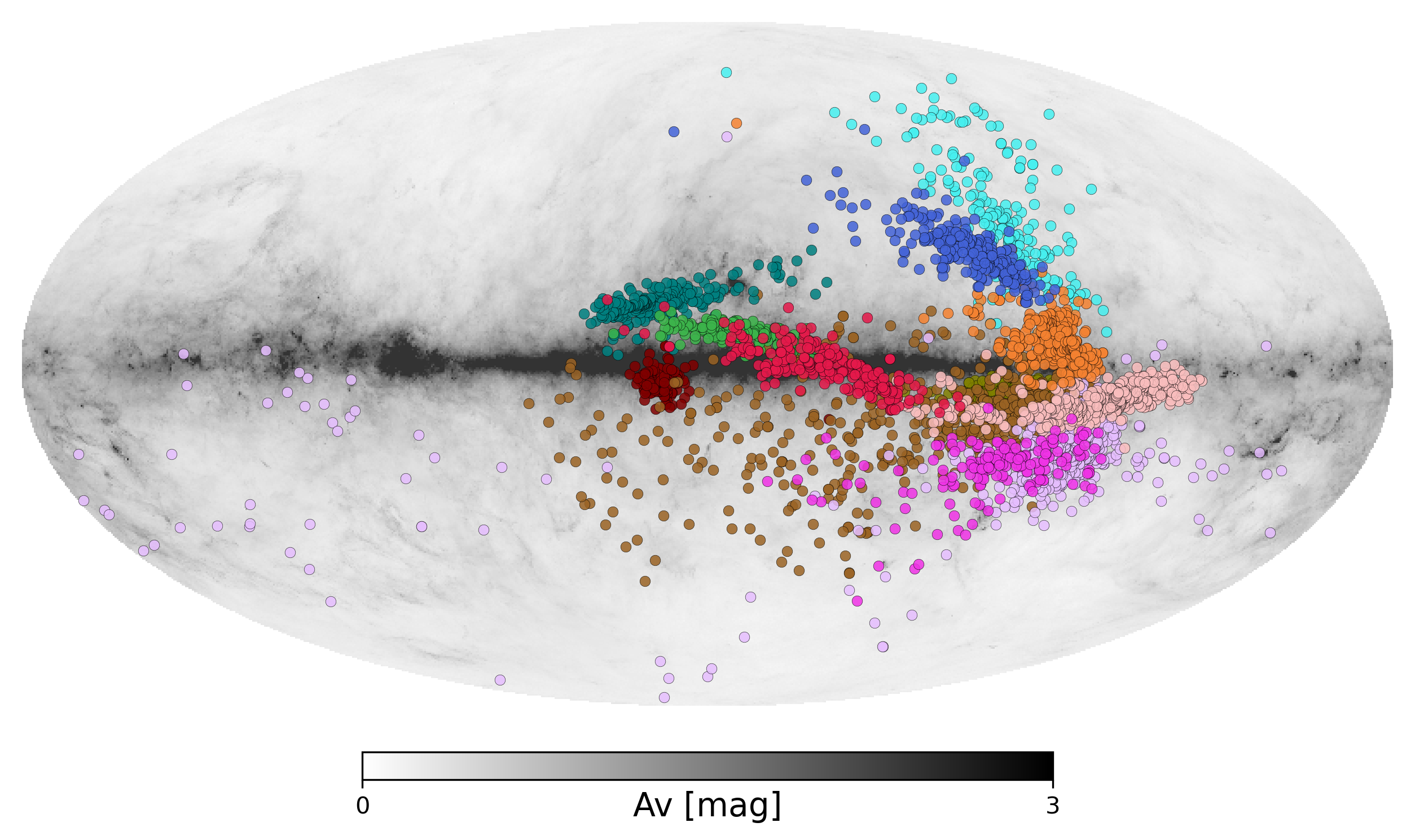}
    \caption{On-sky distribution of our selection for 12 disk streams on top of the \textit{Planck} dust map~\citep{Planck:2013}. All the streams were identified inside a fully sampled 250$^3$ pc$^3$ in the local Milky Way. Colors have the same meaning as in Fig.~\ref{fig:3d_view}.}
    \label{fig:four_pi}
\end{figure*}

\subsection{Age and mass}\label{sect:age_mass}
% ------------- Group stats table -----------------
\begin{table*}
\begin{center}
\begin{small}
\caption{Overview of the computed cluster parameters and statistics of the 12 identified disk streams.\vspace{-1.5mm}}
\renewcommand{\arraystretch}{1.4}
\resizebox{0.995\textwidth}{!}{
\begin{tabular}{llrrrrrcccccr}

\hline \hline

\multicolumn{1}{l}{SID} &
\multicolumn{1}{l}{Literature} &
\multicolumn{1}{c}{Size} &
\multicolumn{1}{c}{Age\,\tablefootmark{g}} &
\multicolumn{1}{c}{Literature age\,\tablefootmark{b}} &
\multicolumn{1}{c}{Reference\,\tablefootmark{c}} &
\multicolumn{1}{c}{Length\,\tablefootmark{d}} &
\multicolumn{1}{c}{Aspect ratio} &
\multicolumn{1}{c}{M$_{\text{tot}}$} & 
\multicolumn{1}{c}{M$_{\text{sys}}$\,\tablefootmark{d}} & 
\multicolumn{1}{c}{$\sigma_{3D}$\,\tablefootmark{h}} & 
\multicolumn{1}{c}{$\Vec{V}$ (\emph{UVW})} &
\multicolumn{1}{c}{S/N} \\[-1mm]
 &
\multicolumn{1}{l}{name\,\tablefootmark{a}} &
 &
\multicolumn{1}{c}{(Myr)} &
\multicolumn{1}{c}{(Myr)} &
 &
\multicolumn{1}{c}{(pc)} &
 &
\multicolumn{1}{c}{($\text{M}_{\odot}$)} & 
\multicolumn{1}{c}{($\text{M}_{\odot}$)} & 
\multicolumn{1}{c}{(km s$^{-1}$)} & 
\multicolumn{1}{c}{(km s$^{-1}$)} &
 \\[0.5mm]
\hline

1 & Ratzenboeck~1 & 273 & $103^{\phantom{0}+19}_{\phantom{00}-7}$ & - & - & $320 \pm 10$ & $\phantom{0}7.5 : 1.3 : 1$ & $140$ & $369 \pm 37$ & 5.1~(7.9) & $(-6.6, -26.9, -11.7)$ & $\phantom{0}10 \pm \phantom{0}2$  \\[1mm]
2 & Theia 368 & 172 & $\phantom{0}96^{\phantom{0}+17}_{\phantom{0}-11}$ & 140, 93, 126, 158 & 2, 3, 4 7 & $120 \pm 20$ & $\phantom{0}7.8 : 2.2 : 1$ & $\phantom{0}96$ & $198 \pm 34$ & 2.8~(3.8) & $(-1.0, -28.9, -14.0)$ & $\phantom{0}76 \pm 16$ \\[1mm]
3 & Theia 430 & 133 & $288^{\phantom{0}+35}_{\phantom{0}-20}$ & 80, 190  & 3, 4 & $250 \pm 20$ & $\phantom{0}7.6 : 1.9 : 1$ & $144$ & $310 \pm 23$ & 2.5~(4.4) & $(-26.4, -7.5, -10.7)$ & $\phantom{0}10 \pm \phantom{0}3$ \\[1mm]
4 & Platais 9 & 454 & $\phantom{0}66^{\phantom{00}+4}_{\phantom{00}-2}$ & 78, 50, 37, 54 & 1, 2, 3, 4 & $280 \pm 30$ & $\phantom{0}6.9 : 1.7 : 1$ & $248$ & $524 \pm 41$ & 2.8~(4.2) & $(-23.6, -15.4, -6.1)$ & $\phantom{0}21 \pm 11$ \\[1mm]
5 & HSC 2278 & 203 & $976^{+251}_{-185}$ & 571 & 3 & $280 \pm 30$ & $\phantom{0}8.2 : 2.6 : 1$ & $\phantom{0}80$ & $168 \pm 20$ & 2.7~(4.3) & $(-4.9, -4.4,\phantom{0}0.1)$ & $\phantom{00}5 \pm \phantom{0}1$ \\[1mm]
6 & Theia 371\,/\,OCSN 87 & 172 & $158^{\phantom{0}+26}_{\phantom{0}-24}$ & 355, 144, 146, 631 & 2, 3, 4, 7 & $300 \pm 30$ & $\phantom{0}5.6 : 1.9 : 1$ & $105$ & $262 \pm 15$ & 2.1~(2.7) & $(-0.6, \phantom{0}1.3, -1.1)$ & $\phantom{0}12 \pm \phantom{0}3$ \\[1mm]
7 & NGC 2451A & 845 & $\phantom{0}49^{\phantom{00}+5}_{\phantom{00}-2}$ & 44, 56, 26, 50 & 1, 2, 3, 4 & $330 \pm 10$ & $\phantom{0}5.3 : 1.6 : 1$ & $435$ & $812 \pm 77$ & 3.5~(5.7) & $(-27.3, -14.4, -12.7)$ & $\phantom{0}28 \pm \phantom{0}9$ \\[1mm]
8 & Mamajek 2 & 324 & $128^{\phantom{0}+13}_{\phantom{0}-18}$ & 126, 99, 178, 125 & 2, 3, 4, 5 & $220 \pm 10$ & $\phantom{0}8.4 : 2.3 : 1$ & $169$ & $380 \pm 25$ & 3.4~(5.9) & $(-10.3, -25.8, -4.4)$ & $\phantom{0}21 \pm \phantom{0}8$ \\[1mm]
9 & Theia 301\,\tablefootmark{e} & 534 & $101^{\phantom{0}+11}_{\phantom{00}-6}$ & 107 & 4  & $430 \pm 10$  & $\phantom{0}7.1 : 2.3 : 1$ & $279$ & $652 \pm 68$ & 2.1~(3.6) & $(-8.9, -27.3, -11.8)$ & $\phantom{00}6 \pm \phantom{0}2$ \\[1mm]
10 & Volans-Carina & 566 & $106^{\phantom{0}+23}_{\phantom{00}-5}$ & 355, 212, 189, 89  & 2, 3, 4, 6 & $280 \pm 10$ & $\phantom{0}6.6 : 1.6 : 1$ & $255$ & $570 \pm 44$ & 3.8~(5.6) & $(-16.2, -27.0, -0.46)$ & $\phantom{00}9 \pm \phantom{0}4$ \\[1mm]
11 & OCSN 3 & 196 & $411^{\phantom{0}+75}_{-109}$ & 282, 705 & 2, 3 & $160 \pm 40$ & $\phantom{0}3.3 : 1.0 : 1$ & $120$ & $328 \pm 26$ & 2.1~(3.3) & $(-7.9, -17.2, -13.4)$ & $114 \pm 21$ \\[1mm]
12 & Theia 599\,\tablefootmark{f} & 418 & $461^{+106}_{\phantom{0}-26}$ & 264 & 4 & $350 \pm 10$ & $10.1 : 2.3 : 1$ & $233$ & $527 \pm 53$ & 3.1~(4.5) & $(-35.5, -17.0, -0.6)$ & $\phantom{0}22 \pm \phantom{0}9$ \\[1mm]
\hline
\end{tabular}
}
\renewcommand{\arraystretch}{1}
\label{tab:group_stats}
\tablefoot{
    \tablefoottext{a}{
    We aim to report the name assigned to a cluster based on its first literature appearance, given a one-to-one relationship and enough similarity between our and the literature selection. For clusters that appear to be contaminated or describe a smaller (noncentral) portion of the disk stream, we also report the name of a more recent study whose extraction aligns better with our findings. This is particularly the case for disk streams S1 and S6, where we have either established a new name (Ratzenboeck~1) or attributed two names to describe the population.
    }
    \tablefoottext{b}{
    Literature ages are compiled from the discovery paper and/or follow-up studies with refined membership lists. In cases of very contaminated extractions, we omit the age altogether (see Table\,\ref{tab:comparison}). For the open clusters Platais 9 and NGC 2451A, we report the age determined by \citet{Bossini2019}; see \citet{Meingast2021} for a more thorough discussion on their fundamental physical parameters across the literature. Here, we report the mean population age rounded to the nearest integer and omit reported uncertainties for easier readability. We refer the reader to the respective source material referenced in the respective ``Reference'' column for more information. 
    }
    \tablefoottext{c}{Ages listed in the ``Literature age'' column correspond to the following references: (1) \citet{Bossini2019}, (2) \cite{Qin:2023}, (3) \citet{Hunt:2023}, (4) \citet{Kounkel2019}, (5) \citet{Mamajek:2006}, (6) \citet{Gagne:2018:VolCar}, (7) \citet{Fuernkranz:2024}
    }
    \tablefoottext{d}{Due to possible contamination, we determine the disk stream lengths not directly from the data but as average peak-to-peak lengths from 100 bootstrap samples, which we report alongside respective standard deviations.}
    \tablefoottext{e}{The disk stream S9 overlaps with the group Theia 301 identified by \citet{Kounkel2019} who relate it to the AB Doradus moving group. We note here that our selection of S9 is significantly different from Theia 301. While Theia 301 consists of just over 1\,300 sources, only 188 of them overlap with S9, which itself has over 500 members. Regardless of these differences, we refer to S9 as ``Theia 301'' throughout this manuscript.}
    \tablefoottext{f}{
    The disk stream S12 overlaps with the group Theia 599 identified by \citet{Kounkel2019}. Their morphologies and membership lists are similar enough to document a match here. However, we note that the observational HRD and \emph{XYZ} distribution of Theia 599 indicates substantial field contamination.
    }
    \tablefoottext{g}{
    We find significantly nonzero (A$_\text{V} \geq 0.1$\,mag) extinction toward disk streams S1 (A$_\text{V} = 0.2$ mag), S3 (A$_\text{V} = 0.1$\,mag), S4 (A$_\text{V} = 0.1$\,mag), S8 (A$_\text{V} = 0.5$\,mag), S11 (A$_\text{V} = 0.3$\,mag), S12 (A$_\text{V} = 0.1$\,mag).
    }
    \tablefoottext{h}{
    We show two estimations of the 3D velocity dispersion. First, the dispersion estimate is derived from the XD procedure. Second, the value in parentheses is determined via the MAD. For more details, see Appendix~\ref{app:vel_disp}. 
    }
}
\end{small}
\end{center}
\end{table*}
% -------------------------------------------------

To estimate the masses and ages of the streams, we adopted the isochrone fitting procedure from\,\citet{Ratzenboeck:2023b} \citep[using PARSEC;][]{Marigo:2017}, where noise contributions around isochronal curves are modeled via skewed Cauchy distributions. The skewed Cauchy distribution naturally accounts for nonsymmetric noise sources, such as unresolved binaries and differential reddening effects inside the cluster. At the same time, its heavy tails are known for their robustness to outliers\,\citep{Hampel:2011}. We derived the total stellar mass (M$_{\text{tot}}$) using the inferred isochronal ages and the sources' relative positions to the best-fitting isochrone (assuming solar metallicity $Z_{\odot} = 0.0152$) in the \emph{Gaia} color--absolute magnitude diagram (using G$_{\text{BP}}$ – G$_{\text{RP}}$ as color). 

Combined cluster masses of the entire system (M$_{\text{sys}}$) are estimated by taking into account \emph{Gaia}'s incompleteness toward very bright and faint objects. Assuming a Kroupa initial mass function\,\citep[IMF; here][]{Kroupa:2001}, we determined M$_{\text{sys}}$ by fitting an IMF to the observed mass distribution as a function of variable cluster mass. Figure~\ref{fig:masses} shows the observed mass functions for the 12 disk streams along with the best-fitting IMF. In practice, for each cluster, we binned ($N$=10) the mass range where \emph{Gaia} is approximately complete and minimized the $\chi^{2}$ statistic (i.e., the normalized sum between the squared difference of observed and estimated counts across all mass bins). This mass range is determined from the G-band range between magnitudes 12 and 17 where \emph{Gaia} DR3 is expected to be complete~\citep{Riello2021}, which we subsequently translated to a mass range using the distances to cluster members to obtain absolute magnitudes and corresponding isochrones. An as yet under-explored field concerns the systematic effects of clustering algorithms on derived physical parameters. For the mass function, we suspect that the cluster selection function is a second-order effect that does not drastically affect our results. However, these selection biases should be considered when studying the detailed shape of the observational mass function.

Table\,\ref{tab:group_stats} summarizes the main physical parameters of the 12 identified disk streams and compares their ages to those reported in the literature. We find that our age estimates, in general, agree with other references. The ages of the disk streams range from about $50$\,Myr to around $1$\,Gyr, with a median age of 117\,Myr. Most streams (8 of 12) are relatively young, with ages less than $200$, Myr. 

We find no apparent relationship between a population's age and length, as measured by the sample Pearson correlation coefficient ($r = -0.06$). This observation contradicts expectations from tidal disruption processes, where stars gradually migrate into a system’s tidal tails. However, as explained above, the lengths derived in this work are lower limits, and this result will likely change with better \emph{Gaia} DR4 data. In contrast, we observe a minimal correlation between a cluster's volume and age ($r = 0.1$), although with considerable scatter. On average, our findings suggest that (when ignoring outliers) the volume of disk streams increases by approximately $500 - 1000$\,pc$^3$ per Myr. Figure~\ref{fig:correlations} provides an overview of all pairwise correlations mentioned in this section.

We find that the disk stream mass can moderately predict the length of a stream ($r = 0.55$), which is reasonable since a larger system mass provides a higher signal contrast over the background for the entire stream size. A related statistic, the population density (stars per cubic parsec) over time, shows a moderate negative correlation ($r = -0.56$). This anticorrelation suggests that populations dissolve into the surrounding field over the lifetime of a cluster.

\subsection{Boundedness estimation}
Similarly to \citet[hereafter \citetalias{Meingast2021}]{Meingast2021}, our objective was to investigate the ``boundedness'' and dissolution process of recovered populations. Compared to their study, we did not start from a set of prominent open clusters but rather aimed to analyze all stream-like structures inside a given search volume. Therefore, we hypothesize that most identified disk streams do not have a bound core. To assess the boundness of a cluster's core, we analyzed the Jacobi radius, $r_J$ (i.e., the dynamical tidal radius) of each cluster population as outlined in \citetalias{Meingast2021} and \citet{Hunt:2024}. To do so, we computed the cumulative total and completeness-corrected mass enclosed at varying distances, M$_{\mathrm{obs}}(r),$ from the density mode of each population, estimated with a kernel density estimate (KDE); for a detailed description, we refer the reader to \citet{Hunt:2024}, whose implementation we have adopted. By intersecting the observed radial mass profile with the theoretical Jacobi mass for a cluster of a given size M$_J(r)$, denoting the distance from a cluster center to the Lagrange points $L_1$ and $L_2$ of a bound system \citep{King:1962, Ernst:2010, PortegiesZwart:2010}, we obtained a system's Jacobi radius, $r_J$, and mass, M$_J$. We note that although the Jacobi radius procedure as implemented \citet{Hunt:2024} (and here) has some differences from its application in \citetalias{Meingast2021}, the Jacobi radius results for NGC 2451A and Platais 9 (both groups appear in \citetalias{Meingast2021} and this study) agree within 0.5\,pc and 2.3\,pc, respectively. 

We employed the definition of \citet{Hunt:2024} to determine whether a system has a bound core. A partially bound cluster has to have a valid Jacobi radius alongside a minimum enclosed mass of M$_J \geq 40\, \mathrm{M}_{\odot}$. If a cluster has M$_{\mathrm{obs}}(r) < \mathrm{M}_{J}(r)$ for all radii $r$, then no valid Jacobi radius can be determined and the disk stream is classified as entirely unbound. Four of the twelve disk streams meet this boundedness criterion: Platais 9, NGC 2451A, Mamajek 2, and OCSN 3. Table~\ref{tab:jacobi} lists their respective Jacobi radii and masses alongside the fraction of bound mass inside the disk stream. As first reported by \citetalias{Meingast2021}, we also find that most mass, on average around $65\%$, of these four groups does not reside in their bound core but rather in the tails or coronae of these disk streams. The remaining disk streams -- Ratzenboeck~1, Theia 386, Theia 430, HSC 2278, Theia 371/OCSN 87, Theia 301, Volans-Carina, and Theia 599 -- appear to be fully unbound.

Lastly, we observe a small negative correlation between the age of the stream and its 3D velocity dispersion. This result is unexpected, as we anticipated an increase in velocity dispersion with age due to processes such as disk heating and GMC encounters. Similarly, the lack of correlation between a stream’s age and its length challenges the assumption that tidal tails grow uniformly over time. These findings are presumably explained by the diversity in the initial properties of the clusters, such as stellar density and mass, alongside the time since disruption and the specifics of the disruption process itself, all of which strongly influence the streams' present day phase space distribution and survival times.
Furthermore, the observed lack of correlation might also suggest limitations in our pipeline’s sensitivity to distinguish the low-density, highly dissolved tails from the surrounding field. These sources could potentially be recovered more effectively by clustering in a different feature space, such as action-angle coordinates \citep[see, e.g.,][]{Fuernkranz:2024}, or through higher-resolution measurements from upcoming \emph{Gaia} data releases.

\subsection{Comparison with established cluster catalogs}\label{sect:comparison}
Most stellar structures we identify can be crossmatched with literature samples. Table\,\ref{tab:comparison} lists the 12 disk streams and provides an overview of the literature matches we find. Our search box also contains two prominent young open star clusters, NGC 2451A and Platais 9, around which we find substantial coronae, previously studied in detail by \citetalias{Meingast2021}. Our pipeline reproduces their results in cluster morphology and approximate size. 

An in-depth comparison is provided in Appendix~\ref{app:comparison}, where we also compare our results across multiple literature catalogs\footnote{Cluster papers not mentioned in this section or Table\,\ref{tab:comparison} do not crossmatch to the identified disk streams or are already comprehensively discussed by \citetalias{Hunt:2023}; we refer the reader to this work for further in-depth matches with catalogs not covered here.}. Here in the main text, we focus on the cluster catalog of \citet{Hunt:2023} (hereafter, \citetalias{Hunt:2023}), which represents an unsupervised and homogeneous state-of-the-art reference that exhibits a high level of similarity in cluster morphology compared to our selection. Out of the 12 disk streams identified, 9 groups have clear counterparts in \citetalias{Hunt:2023}. On average, we identify twice as many sources compared to these counterparts.

% -------- Jacobi table -------------
\begin{table}
\begin{center}
\begin{small}
\caption{Overview of the Jacobi radii, $r_J$, and masses, M$_J$, alongside the bound mass fraction of the four clusters where we find a bound core; see Sect.~\ref{sect:age_mass} for more information.\vspace{-1,5mm}}
\renewcommand{\arraystretch}{1.1}
\resizebox{0.87\columnwidth}{!}{
\begin{tabular}{llccc}

\hline \hline \\[-3.3mm]

\multicolumn{1}{l}{SID} &
\multicolumn{1}{l}{Literature} &
\multicolumn{1}{c}{$r_J$} &
\multicolumn{1}{c}{M$_J$\,\tablefootmark{a}} &
\multicolumn{1}{c}{Bound mass fraction\,\tablefootmark{b}} \\[-0.5mm]

 &
\multicolumn{1}{l}{name} &
\multicolumn{1}{c}{(pc)} &
\multicolumn{1}{c}{(M$_{\odot}$)} &
\multicolumn{1}{c}{(\%)} \\[0.5mm]
\hline\\[-3mm]

4 & Platais 9 & 5.9 & 79.5 & 32.1 \\[1mm]
7 & NGC 2451A & 7.7 & 170.2 & 39.1 \\[1mm]
8 & Mamajek 2 & 5.1 & 56.0 & 33.2 \\[1mm]
11 & OCSN 3 & 4.6 & 44.2 & 36.9 \\[1mm]

\hline
\end{tabular}
}
\renewcommand{\arraystretch}{1}
\label{tab:jacobi}
\tablefoot{
    \tablefoottext{a}{
    The Jacobi mass is determined using the observed masses (see M$_{\text{tot}}$) and not the system masses that are derived from fitting a Kroupa IMF to the observed mass histograms.
    }
    \tablefoottext{b}{
    The boundedness fraction is determined by dividing the Jacobi masses by the observed system masses, i.e.,  M$_{\text{tot}}$ as provided in Table~\ref{tab:group_stats}; see also footnote (a).
    }
}
\end{small}
\end{center}
\end{table}

\begin{figure}
    \centering
    \includegraphics[width=0.7\columnwidth]{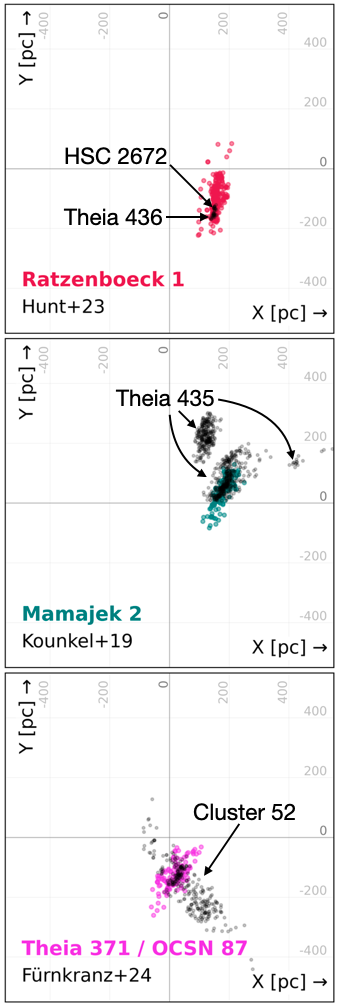}
    \caption{Examples of challenging cluster comparisons highlighted in the \emph{X-Y} plane. The colored scatter points show the identified disk streams Ratzenboeck~1, Mamajek 2, and Theia 371/OCSN 87 (from top to bottom). The black scatter points show the crossmatch to a literature cluster in the \citetalias{Hunt:2023}, \citetalias{Kounkel2019}, and \citet{Fuernkranz:2024} surveys (from top to bottom). See Sect.\,\ref{sect:comparison}, Appendix\,\ref{app:comparison}, and Table\,\ref{tab:comparison} for further details.}
    \label{fig:cluster_comparison}
\end{figure}

When comparing our results across cluster catalogs in the literature, it appears that disk stream S1 has not yet been identified as such in previous research. Our selection of S1 shows minimal overlap with \citetalias{Hunt:2023}, who detected two small fragments, each constituting approximately 5\% and 7\% of S1 (see Fig.\,\ref{fig:cluster_comparison}, top panel). Hence (as briefly mentioned in Sect.~\ref{sect:uncover}), we have named the population ``Ratzenboeck~1.''

Comparisons with other catalogs are more challenging due to sometimes differing cluster morphologies in the literature, which complicates direct comparisons. Figure~\ref{fig:cluster_comparison} illustrates these challenges. The middle and bottom panels show comparisons with \citet{Kounkel2019} and \citet{Fuernkranz:2024}, respectively. While some cluster sources align, the overall cluster distributions can appear significantly different.
% ---------------------

\begin{figure*}
    \centering
    \includegraphics[width=0.95\textwidth]{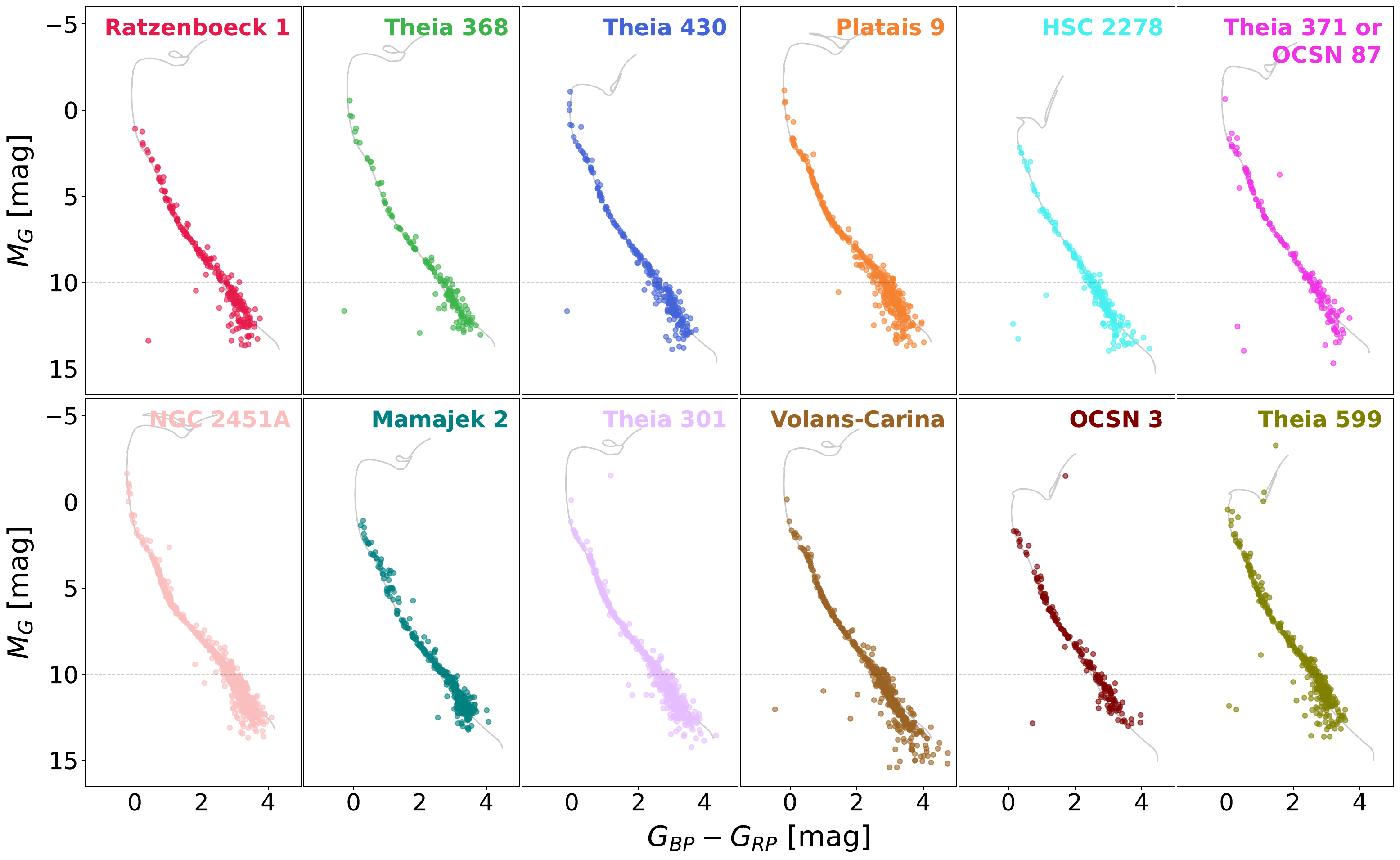}
    \caption{Observational HRDs for the 12 identified disk streams. Sources that do not satisfy the astrometric quality criterion RUWE < 1.4 \citep{Lindegren2021} or have large photometric uncertainties ($G_{\text{err}}$ < 0.007 mag; $G_{\text{RP,err}} < 0.03$ mag; $G_{\text{BP,err}} < 0.15$ mag) have been removed to reduce large random scatters due to bad measurements. Colored points represent each cluster's members. The corresponding best-fitting isochrones are shown as gray lines; their respective ages can be found in Table\,\ref{tab:group_stats}. The dashed horizontal line indicates the fitting range limit. Sources fainter than absolute G magnitudes of 10 are removed from the fit due to empirical discrepancies between isochronal curves and data points\,\citep[see][]{Ratzenboeck:2023b}.}
    \label{fig:hrd}
\end{figure*}

Lastly, we briefly highlight the potential relationship between several well-known clusters and associations and our disk stream. First, our analysis indicates a possible overlap between the Coma-Berenices cluster (also known as Melotte 111; using the selection by\,\citet{Fuernkranz2019} who identify its tidal tails) and the group HSC 2278, suggesting that HSC 2278 is the trailing (tidal) tail of the Coma-Ber cluster. 

Second, we find a potential relationship between our selection of Theia 301 (S9) and the AB Doradus moving group\,\citep[AB Dor;][]{Zuckerman:2004} when comparing it to the AB Dor selection of\,\citet{Gagne2018c}. As proposed by\,\citet{Gagne:2021}, AB Dor and Theia 301 may be parts of the same system (alongside the Pleiades and other Theia members). In both cases, the analysis of both groups' space motion and observational HRDs reveals neither definitive support nor a contradiction to the claim of a single-coeval population, and further data are required to generate a definitive answer. 

Third, we find 47 crossmatches between the X-ray-selected Sco-Cen members of \citet{Schmitt2022} and our disk streams. We investigated these candidates in the observational HRD and find that these sources are clearly separated from the young pre--main sequence stars of Sco-Cen, pointing toward a potential source of contamination in X-ray-only selections of young sources. 

\section{Discussion}
Recent \textit{Gaia} data releases have enabled researchers to discover several stream-like structures in the local Milky Way. Beyond the fiducial disk stream Meingast-1, tidal tails have been detected around open clusters \citep[see][]{MeingastAlves2019, Roeser2019} and are now a ubiquitous feature around open clusters, reaching almost 100 detections~\cite{Tarricq:2022}. More recently, cluster coronae \citep[see][]{Meingast2021, Moranta:2022} have been detected as a loose coeval ensemble of stars surrounding dense cluster cores that are likely also produced in tidal disruption processes. However, it is not clear at the moment the impact of the original gas distribution or the relative roles of residual gas expulsion and violent relaxation (see \citetalias{Meingast2021}). Beyond that, further low-density and elongated structures have been identified, such as stellar snakes \citep{Tian2020-bv}, filamentary structures \citep{Beccari:2019}, and the stellar disk stream Theia 456 \citep{Kounkel2019, Andrews:2022}. 

We note that the physical difference between these structures is not clear (yet), and various names might be used to refer to similar dissolution processes. This study does not focus on the origins of disk streams or similar structures, and we call for a more comprehensive sample for an in-depth analysis of their origin. Nevertheless, our preliminary findings warrant a brief discussion. Akin to Meingast-1 and cluster coronae (see \citealt{Meingast2019} and \citetalias{Meingast2021}), the populations identified in this work show similar patterns, such as large elongations $> 100$\,pc, and inclination angles in the \emph{X-Y} plane reminiscent of Galactic tidal interactions. Tidal disruption processes are a plausible explanation for most populations that appear to have a symmetrical leading and trailing arm oriented toward and away from the Galactic center.

\subsection{Peculiar phase space signatures}\label{sect:discussion_vel_disp}
Notably, velocity dispersions in Galactic Cartesian velocity space below $\sim 5$\,km\,s$^{-1}$ are exceptionally low for structures extending over several hundred parsecs. In this space, Galactic rotation has a contribution to the total velocity dispersion for such large structures, which ranges from 1 -- 3\,km\,s$^{-1}$. Appendix~\ref{app:vel_disp} discusses this issue in detail, while we summarize two major points in the following. One potential explanation for these low values is the nature of the deconvolution process. To provide a more robust estimate and minimize the potential bias introduced by the deconvolution method, we also computed velocity dispersions using the MAD (see Sect.~\ref{sect:result_vel}). This robust measure yields slightly higher values, with velocity dispersions ranging from 2.7 -- 7.9\,km\,s$^{-1}$ (see Table~\ref{tab:group_stats}). Another likely source of underestimation arises from selection effects inherent in density-based clustering methods. Given the large extent of the identified structures, the recovered members are embedded within a significant background population. Members in the tails of the velocity distribution, just a few km,s$^{-1}$ from the central overdensity (in any coordinate system), lack sufficient contrast relative to the dominant background (i.e., a S/N of $\sim 1$). As a result, these members often go undetected by clustering algorithms. Addressing this limitation by identifying and recovering the ``missing'' members with higher velocity dispersions is outside the scope of this work but represents a key direction for future work.

Beyond velocity space, the positional distribution raises an important question regarding the shared origins of several structures. As discussed in Sect.~\ref{sect:results}, several disk streams exhibit significant overlaps in position space. However, a detailed kinematic analysis reveals that, in all but one case, these co-spatial streams show statistically significant differences in their 3D velocities and ages. In the case of Ratzenboeck~1 and Theia 386, we find similar 3D velocities and ages, which suggests they represent fragments of a larger structure or a joint formation scenario akin to substructures identified in various OB associations \citep[e.g.,][]{Damiani2019, Chen2020, Kerr2021, Ratzenboeck:2023a}. A traceback analysis of their Galactic orbits in the past and future 20\,Myr alongside a distinct bi-modal signal in their joint phase space distribution (see Appendix~\ref{app:kinematic_pop_analysis}) supports their classification as separate clusters. For a comprehensive discussion of the pairwise kinematic comparisons, including the Mahalanobis distances quantifying these differences and the implications for stream independence (see Appendix~\ref{app:kinematic_pop_analysis}). We also refer readers to \citet{Fuernkranz2019} for an earlier analysis of co-spatial populations and their potential connections.

\subsection{Estimating the disk stream number density}
Using detailed N-body simulations (see \citealt{Kamdar:2019_sim} and \citealt{Kamdar:2021_stream}\footnote{We note that this particular research article has not yet been officially published. However, its results build on disk streams arising from N-body simulations of the Galaxy \citep[see][]{Kamdar:2019_sim} that are independent of this particular research and have been successfully peer-reviewed and published.}, hereafter \citetalias{Kamdar:2021_stream}) recently estimated the number of disk streams we can find considering the capabilities of \emph{Gaia} DR2. They considered two major factors that restrict their detection. First, limited accuracy of astrometric measurements, which restricts the identification of populations with low-density contrast (against the background). And second, destructive encounters with GMCs. The authors tested the impact of different initial conditions in which the progenitor star cluster is born, such as initial cluster mass and dynamical state. They conclude that with the astrometric precision of \emph{Gaia} DR2, 1 to 10 disk streams are likely yet to be found in the solar neighborhood (defined by the authors as a 500\,pc radius sphere centered on the Sun). This number, the authors argue, will improve by a factor of 5 -- 10 with \emph{Gaia} 10-year end-of-mission data where high-fidelity parallaxes allow an increase in the effective search volume out to a radius of 1.5 kpc. These N-body estimates suggest that \emph{Gaia} data are and will continue to be sensitive enough to reach volume densities of around 2 to 20 disk streams per kpc$^3$. 

As discussed above, many elongated, stream-like structures have been identified in the Milky Way disk. In this work, we aim to provide the first empirical estimate of the abundance and, more specifically, the number density of dynamically cold, coeval stream-like structures in our Galaxy. Taking a volume-controlled sample, we observe an abundance of stream-like structures within our local test volume of $250^3$\,pc$^3$. For this volume, which samples $|Z| < 100$, we derive a density of approximately 820 disk streams per kpc$^3$. This estimate is one to two orders of magnitude higher than predicted by N-body simulations. The equivalent surface density estimate on the Galactic \emph{X-Y} plane is around 160 objects per kpc$^{2}$. The co-occurrence of 12 stream-like structures within such a confined region creates tension with the conventional understanding of these structures, which are believed to be heavily suppressed by destructive interactions with one or a few GMCs \citep[e.g.,][]{Gieles:2006, Kamdar:2021_stream}.

To determine whether the over-classification of disk streams has artificially inflated the estimated disk stream density, we evaluated the impact of refining the classification criteria on the density estimates. Specifically, we excluded known open clusters with distinct coronae, such as Platais 9, NGC 2451A, Mamajek 2, and OCSN 3. These clusters feature bound cores, as indicated by our computations (see Sect.~\ref{sect:age_mass}). However, even after these exclusions, the volume density of the remaining disk streams remains an order of magnitude higher than the most optimistic predictions from N-body simulations. This significant discrepancy warrants further discussion on its origin and points either toward a more efficient production mechanism of low-density stream-like disk populations or a less efficient destruction mechanism with a lower disruption time via, for example, GMC interactions \citep[see, e.g.,][]{Kruijssen:2011, Krumholz:2019, Kamdar:2021_stream}. Whichever mechanism dominates, leading to this increased disk stream density, is beyond the scope of this discovery work but will be the subject of future study.

We acknowledge that the proximity of our search box also plays a key role in ensuring data quality, as the precision of astrometric measurements and the resulting phase-space resolution decrease significantly at larger distances. The selected box lies within a region of the Milky Way disk that we believe is representative of typical stellar populations, as supported by its relatively large age spread that approximately matches results from all-sky cluster searches~\citep[e.g.,][]{Hunt:2023, Hunt:2024}. While this region provides high-quality data, we anticipate that similar structures and, thus, similar disk stream densities will be detected in other regions of the Milky Way disk.

In future work, we plan to extend this analysis by repositioning the box to various regions of the Galactic disk. This will increase the currently limited statistical sample of disk streams, allowing us to assess whether our local measurements are typical and to better understand the processes underlying the formation and survival of these structures. Having access to a larger sample of disk streams will also open the door to deriving more detailed constraints on cluster disruption times along GMC properties, such as their mass function, number density, and overall formation and evolution in the Milky Way.

\subsection{Cluster dissolution}\label{sect:dissolution}

\citetalias{Kamdar:2021_stream} find four main predictors for star clusters to eventually become detectable streams. Those are (a) large initial cluster masses, (b) ``young'' ages below 1\,Gyr, (c) the number and severity of GMC interactions, and (d) an initial dynamical state that is preferentially bound, having initial virial ratios of less than one half. While a higher initial cluster mass, a younger age, and less disruption from GMCs seem like straightforward predictors, the authors also argue about the importance of initial boundedness, which enhances a stream's chance of being detected in the present day. 

\begin{figure}
    \centering
    \includegraphics[width=0.99\columnwidth]{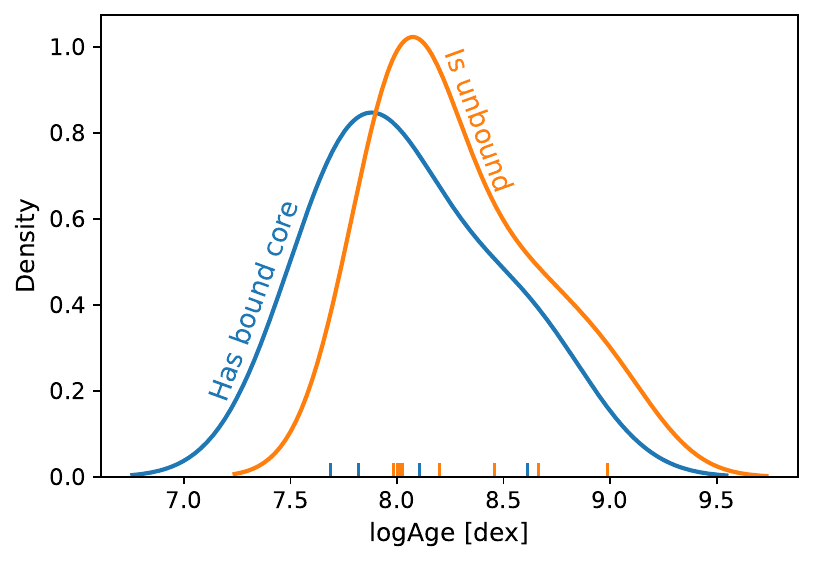}
    \caption{KDE (colored lines) of the age distribution along individual data points (vertical marks on the x-axis), stratified by cluster boundedness: those with evidence of a bound core (blue) and those that are fully unbound (orange). The unbound clusters have a slightly older average age of approximately 100\,Myr. Despite this observed difference, the limited sample size prevents us from drawing statistically significant conclusions regarding the age distributions of the two subpopulations.}
    \label{fig:boundness_vs_age}
\end{figure}

Here, we aim to investigate the role of (partial) boundedness in our sample of 12 disk streams in the context of their age. Using the dichotomy of ``fully unbound'' and ``has a bound core'' as introduced in Sect.~\ref{sect:age_mass}, we searched for any differences in the age distribution between these classes. Figure~\ref{fig:boundness_vs_age} presents the results of this analysis, using KDE\footnote{We estimate the bandwidth of the Gaussian kernel using Scott's rule~\citep{Scott:1979}.} to visualize the age distribution of two distinct subpopulations: clusters with evidence of a bound core (blue) and those that are fully unbound (orange). The individual data points, marked along the x-axis, further illustrate the age distribution within each subpopulation.

Our analysis reveals that fully unbound clusters tend to be, on average, approximately 100\ Myr older than their counterparts with a bound core. Additionally, all but one disk streams beyond 200\,Myr are unbound, except OCSN 3. This observation points to a possible link between cluster age and the likelihood of core dissolution, suggesting that some older (unbound) disk streams we see in our sample initially come from stellar systems that previously had a bound core. This bound core also brings higher survivability of disk streams into old age, as found in N-body simulations, as it makes them more resilient to disruption processes, such as GMC collisions and other Galactic potential variations. At some point, these disruption processes critically accumulate, resulting in fully unbound systems, which are then rapidly dissolved by future GMC interactions and the Galactic tidal field, explaining the low density of disk streams at ages beyond 200\,Myr.

However, it is important to note that the limited sample size prevents us from making statistically significant conclusions regarding the differences in age distributions between the two subpopulations. While the trend is intriguing, more data are needed to robustly determine whether a significant age difference exists between stream-like stellar systems with a bound core and fully unbound disk streams. Another caveat is our simple model of cluster dissolution, ignoring different evolutionary phases and subclasses of stream-like structures such as tidal tails, young moving groups, young local associations, and cluster coronae. In future papers of this series, we aim to work toward a statistically significant sample of disk streams to facilitate the analysis of their different and complex formation and dissolution processes, from Galactic tidal forces, differential rotation, initial star-forming gas configurations, and GMC interactions.

\subsection{Relationship with Sco-Cen}\label{sect:scocen}

The initial sample selection (see Sect.\,\ref{sect:data}) entails that identified disk steams are close to the Sco-Cen OB association. Spatially, the disk streams are tightly packed and can be divided into two groups based on their relative position to Sco-Cen. The disk streams Ratzenboeck~1, Theia 368, Mamajek 2, and OSCN 3 are situated on the far side of Sco-Cen, while the remaining disk streams fill the space between the Sun and Sco-Cen. Remarkably, Ratzenboeck~1, Theia 368, and Mamajek 2 share the same volume to various degrees, with stars associated with the Sco-Cen association. Ratzenboeck~1 appears to ``flow'' through the Sco-Cen subgroups V1062 Sco, and $\mu$\,Sco while partially overlapping with the Cen-Far group (see \citetalias{Ratzenboeck:2023a} for Sco-Cen subcluster names). The disk stream Theia 368 lies almost entirely inside the Sco-Cen subgroups of $\eta$\,Lup, Sco Body, and $\theta$\,Oph. Lastly, although the bulk of Mamajek 2 is outside Sco-Cen, several sources extend into the classical Blaauw subgroup Upper Scorpius.

\begin{figure}
    \centering
    \includegraphics[width=\columnwidth]{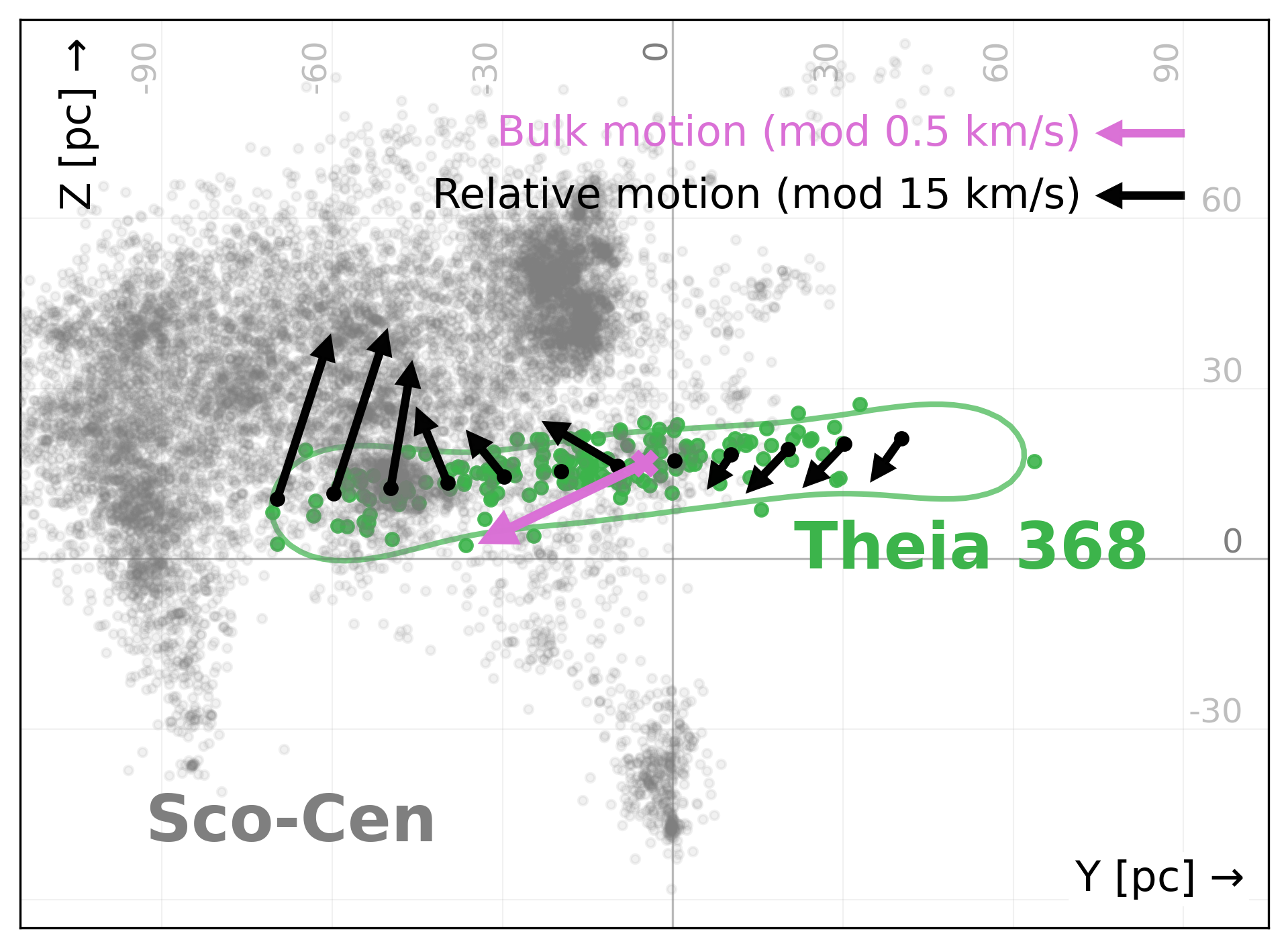}
    \caption{Evidence suggesting an interaction between disk stream Theia 368 (in green) and Sco-Cen (in gray). The black arrows show a running median of the stream's motion relative to its bulk motion, depicted by the purple arrow. The relative motion is scaled up by a factor of 30 compared to the bulk motion to highlight their distribution properly; see the legend on the top left for a size comparison. The relative motions across Theia 368 highly correlate with the stream's interaction time with Sco-Cen; i.e., the further inside Theia 368's members are found in Sco-Cen, the more drastically their motions have been altered.}
    \label{fig:s2_scocen}
\end{figure}

Sco-Cen may have exerted measurable forces on the disk stream Theia 368, which is fully embedded within Sco-Cen. Given that Sco-Cen is located at the edge of the Local Bubble\,\citep{Zucker:2022_localBubble}, which contributes additional gas mass to Sco-Cen's progenitor cloud, there has been sufficient primordial gas mass present to directly and cumulatively influence the disk stream over the past few million years. To determine if this interaction has left measurable effects in present-day observables, we investigated potential imprints in the velocity distribution of Theia 368 members along the length of the stream.

Specifically, we aim to compare the velocity distribution of sources currently embedded within the Sco-Cen association to those currently outside. Figure~\ref{fig:s2_scocen} depicts the potential interaction with Sco-Cen (its members are shown as gray points in the background) that affects the relative 3D velocities observed. The black arrows indicate the motion relative to the stream's bulk motion (purple arrow; see also Table\,\ref{tab:group_stats}) across the entire stream (represented in green). We examined the \emph{Y}-\emph{Z} velocity space to detect this interaction, as it is less influenced by large radial velocities that primarily align with the \emph{X}-axis. To further limit the influence from outliers in \emph{UVW} on the relative velocity signal, we removed sources with radial velocity uncertainties larger than $5\,$km\,s$^{-1}$ and sources flagged as potential outliers by the extreme deconvolution (XD) process (see Appendix~\ref{app:vel_disp}). This quality cut results in a total of 33 sources with ``good'' radial velocity measurements. Lastly, we computed a running median (across five neighbors) at 12 grid points along the \emph{Y}-axis spaced 10\,pc apart (black scatter points in Fig.~\ref{fig:s2_scocen}) to reduce the influence of the remaining random scatter\footnote{We find that binning the data along the \emph{Y}-axis produces a similar pattern. However, some bins contain only one or two members, leading to a significantly greater scatter than a running median.}. 

Sources located outside Sco-Cen (with $Y$\,>\,0\,pc) exhibit nearly constant relative motions that are approximately parallel to the bulk motion (purple arrow in Fig.~\ref{fig:s2_scocen}). In contrast, sources "entering" Sco-Cen show relative velocity vectors that appear to be deflected or scattered by their interaction with the OB association. Moreover, this deflection appears to correlate with the ``depth'' of Theia 368 within Sco-Cen, which gradually increases. These observations provide qualitative and tentative evidence of a scattering process and/or exerted force on parts of the disk stream Theia 368. To substantiate these tentative results, we aim to further investigate this interaction in future work by taking into account a traceback analysis of the unaffected and (apparent) deflected portion of Theia 368 in relation to Sco-Cen, considering the available gas mass and performing a detailed momentum analysis, and contrasting these claims with simpler models that involve only the Galactic potential.

In future work, we aim to investigate similar effects on a larger sample of disk streams, in particular, where our analysis has not yet revealed a clear signal.

\section{Conclusion}
We refined the initial selection of a sample comprising 12 stream-like stellar populations, previously identified as interlopers in \citetalias{Ratzenboeck:2023a}, by using the established machine learning pipelines \texttt{SigMA} and \texttt{Uncover} to improve their census. One disk stream had not been identified in the previous literature, and we have named it Ratzenboeck~1. Compared to previous unsupervised studies, we find, on average, twice as many candidate members per stream. Our pipeline is sensitive to median stream densities (stream densities are averages throughout the volume of the population) of one star per 10$^3$\,pc$^3$ (0.001 stars/pc$^3$), which is about 50 times lower than the surrounding field. At the very extreme, we can recover streams with average densities as low as 0.2 sources per 10$^3$\,pc$^3$ (0.0002 stars/pc$^3$) across the entire population (i.e., resolving structures 250 times below the field density, in \emph{XYZ}).

\begin{enumerate}
    \item The 12 disk streams found within the 250$^3$ pc$^3$ volume yield an estimated volume density of approximately 820~objects\,/\,kpc$^3$ and a surface density of roughly 160~objects\,/\,kpc$^2$. These estimates exceed previous number density calculations by one to two orders of magnitude, as documented by \citetalias{Kamdar:2021_stream}. 
    
    \item We find evidence that the disk stream Theia 368 (S2), predominantly embedded within Sco-Cen, has recently undergone disruption, likely due to interactions with the primordial gas of the OB association.

    \item These 12 disk streams are highly prolate and have lengths between $120$ and  $430$\,pc, with large aspect ratios (longest principal axis over shortest axis) of $3 - 11$.
    
    \item The identified disk streams are dynamically cold, having low 3D velocity dispersions ranging from $2 - 5$\,km\,s$^{-1}$, and show clearly defined and narrow sequences in the HRD, strongly suggesting a coeval nature. Stream ages range from 50\,Myr to 1\,Gyr, with a median age of around 100\,Myr. 
    
    \item We find that disk streams with bound cores are typically younger than fully unbound ones, with fully unbound systems being, on average, about 100\,Myr older. Beyond 200\,Myr, most disk streams are fully unbound, likely reflecting the cumulative effects of disruptive processes such as GMC interactions and Galactic tidal forces.

    \item We have identified an approximately linear relationship between stream ages and their respective volumes, which increase by around 500 - 1\,000 pc$^3$/Myr.
\end{enumerate}
 
\noindent Much like their halo counterparts, disk streams can serve as critical probes for understanding the mass distribution of the Galaxy on both large and small scales. The prevalence of these stream-like features within such a confined region represents a significant departure from conventional understanding, calling for a revision of the formation and dissolution scenarios and for a larger systematic census of disk streams.

\begin{acknowledgements}
    S.\,Ratzenb{\"o}ck acknowledges funding by the Federal Ministry Republic of Austria for Climate Action, Environment, Energy, Mobility, Innovation and Technology (BMK, \url{https://www.bmk.gv.at/}) and the Austrian Research Promotion Agency (FFG, \url{https://www.ffg.at/}) under project number FO999892674. Co-funded by the European Union (ERC, ISM-FLOW, 101055318). Views and opinions expressed are, however, those of the author(s) only and do not necessarily reflect those of the European Union or the European Research Council. Neither the European Union nor the granting authority can be held responsible for them.
    This work has made use of data from the European Space Agency (ESA) mission {\it Gaia} (\url{https://www.cosmos.esa.int/gaia}), processed by the {\it Gaia} Data Processing and Analysis Consortium (DPAC, \url{https://www.cosmos.esa.int/web/gaia/dpac/consortium}). Funding for the DPAC has been provided by national institutions, in particular, the institutions participating in the {\it Gaia} Multilateral Agreement. 
    This research has used Python, \url{https://www.python.org}; \textit{Astropy}, a community-developed core Python package for Astronomy \citep{Astropy2013, Astropy2018}; 
    NumPy \citep{Walt2011}; 
    Matplotlib \citep{Hunter2007}; 
    and Plotly \citep{plotly2015}. 
    This research has made use of the SIMBAD database operated at CDS, Strasbourg, France \citep{Wenger2000}; of the VizieR catalog access tool, CDS, Strasbourg, France \citep{Ochsenbein2000}; and of ``Aladin sky atlas'' developed at CDS, Strasbourg Observatory, France \citep{Bonnarel2000, Boch2014}. This research has used TOPCAT, an interactive graphical viewer and editor for tabular data \citep{Taylor2005}.
\end{acknowledgements}

%-------------------------------------------------------------------
%--------------------------------------------------------------------
\bibliographystyle{aa.bst}

\bibliography{references}

%--------------------------------------------------------------------
% APPENDIX
%--------------------------------------------------------------------
\begin{appendix}
\section{\texttt{SigMA} parameter selection}\label{app:sigma_parameters}
The \texttt{SigMA} pipeline requires the selection of multiple parameters, each contributing to a unique output for a given set of input parameters. Generally, we adopted the parameter choices outlined in \citetalias{Ratzenboeck:2023a}, except for the scaling factor values, which need to be adjusted due to the change in input space from 5D to 6D phase space. As explained in \citetalias{Ratzenboeck:2023a} (see Sect. 3.3.3), the purpose of the scaling factors is to normalize the different data ranges across various subspaces. In our study, the input data space comprises three positional axes (\emph{XYZ}) and three velocity axes (\emph{UVW}). Within each subspace, the Euclidean norm effectively represents the similarity between sources, such as the distance between stars in parsecs or the velocity difference in km\,s$^{-1}$. Generally, our goal would be to normalize one subspace relative to another (see \citetalias{Ratzenboeck:2023a}) based on the characteristic dispersion of the objects we intend to cluster within each subspace. However, stellar streams are notably elongated and exhibit significantly different extents along their three principal axes in \emph{XYZ} space (see Table\,\ref{tab:group_stats}).

Instead of using a single value or a small range of values for the scale factor, we aim to explore a broad space of theoretically meaningful scaling factor values and track a single cluster through this series of \texttt{SigMA} runs. We used the progenitor streams as a reference to determine this range. For each progenitor stream, we calculated the three eigenvalues of its covariance matrix by performing principal component analysis  in both positional and kinematic subspaces. Using these six dispersion coefficients (three for \emph{XYZ} and three for \emph{UVW}), we generated nine scaling factors by considering all possible pairs of these coefficients. We repeated this procedure for all 12 disk streams, generating a distribution of possible scale factors. To exclude extreme outliers, we took this distribution's 5th and 95th percentiles and divided the resulting range into 20 equally spaced scaling factors\footnote{This results in the following scale factors (rounded to the first decimal) with which we multiplied the velocity subspace axes: \{1.5,  3.3,  5.0,  6.8,  8.6, 10.4, 12.1, 13.9, 15.7, 17.5, 19.2, 21.0, 22.8, 24.6, 26.3, 28.1, 29.9, 31.7, 33.4, 35.2\}.}. Running \texttt{SigMA} 20 times yields an ensemble of clustering results in which a given progenitor stream is recovered slightly differently in each run. On average, a stream is identified as such in approximately $35\%$ of the runs, predominantly for runs with lower scale factors below 10. 

After automatically identifying the stream by crossmatching with the progenitor stream in each run, we retained a source as a stream member if it appears in at least $50\%$ of the runs where the stream was successfully recovered. This clustering result is then used to train further the membership pipeline \texttt{Uncover}, as described in Sect.\,\ref{sect:uncover}.

\section{Result supplements}\label{app:mass}
\begin{figure*}
    \centering
    \includegraphics[width=0.99\textwidth]{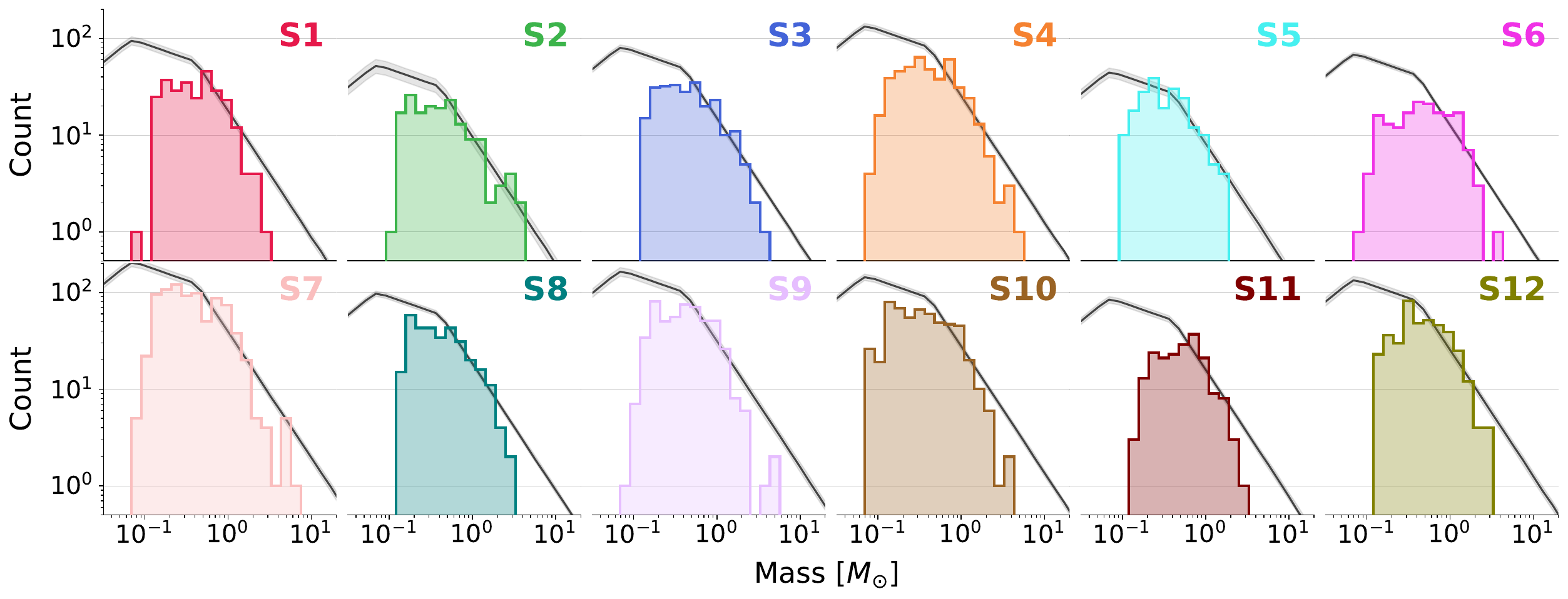}
    \caption{Mass functions for the 12 disk streams used to derive system masses for each population. The best-fitting Kroupa IMFs \citep[solid black line;][]{Kroupa:2001} and $1\sigma$ uncertainties (gray shaded area) are plotted on top of each histogram.}
    \label{fig:masses}
\end{figure*}

This section provides auxiliary information to our result section in Sect.~\ref{sect:results}. Figure~\ref{fig:masses} shows the mass histograms for the 12 disk streams characterized in this work alongside the best fit Kroupa IMFs \citep{Kroupa:2001} and the corresponding uncertainties determined via bootstrap samples 100 times (see  Table~\ref{tab:group_stats} for an overview of all fit values). 

\begin{figure*}
    \centering
    \includegraphics[width=0.85\textwidth]{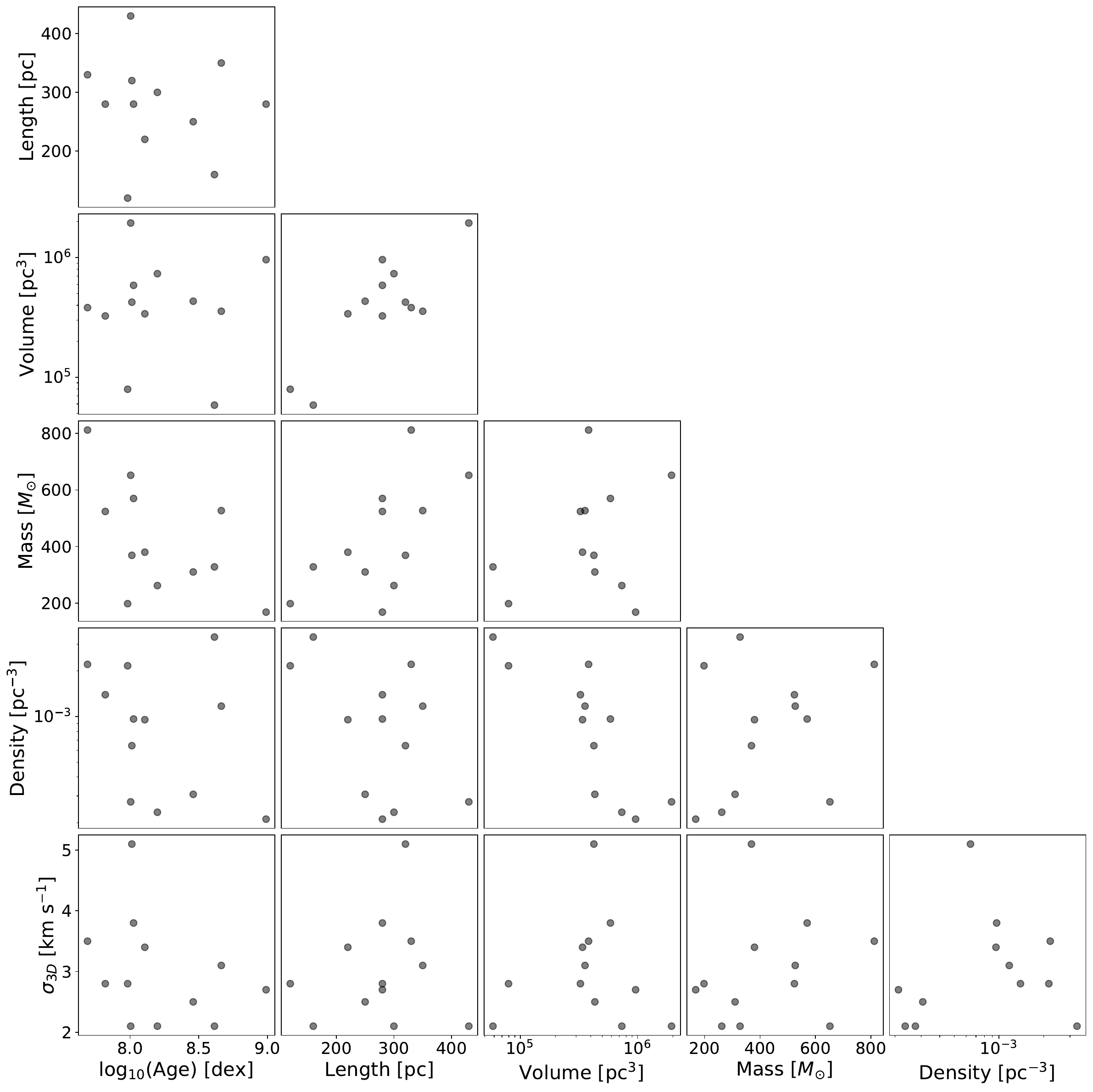}
    \caption{Scatter plot matrix showing pairwise correlations between disk stream ages, lengths, volumes, masses, densities, and 3D velocity dispersions. }
    \label{fig:correlations}
\end{figure*}

Figure~\ref{fig:correlations} is a scatter plot matrix that highlights pairwise correlations among various physical parameters determined for the 12 disk streams. Specifically, Fig.~\ref{fig:correlations} shows the relationships between age, length, volume, mass, density, and 3D velocity dispersion for all disk streams (see   Table~\ref{tab:group_stats} for an overview of the physical parameters). 

\section{Literature comparison supplements}\label{app:comparison}
This section provides auxiliary information to our literature comparison presented in Sect.\,\ref{sect:comparison}. Table\,\ref{tab:comparison} overviews the 12 identified disk steams and their matches in the literature. 

Our comparison of the literature reveals that we find the best agreement in stream morphology and approximate size compared to the work of \citet[hereafter \citetalias{Hunt:2023}]{Hunt:2023}. Seven of the 12 identified disk streams have a similar counterpart in \citetalias{Hunt:2023} (S2, S3, S4, S6, S7, S8, and S10). In addition, groups S5 and S11 also have a clear counterpart in \citetalias{Hunt:2023}, albeit with a fairly smaller extent and source count. The streams S6 and S11, named OCSN 87 and OCSN 3, respectively, were first identified by \citet[hereafter \citetalias{Qin:2023}]{Qin:2023}. \citetalias{Qin:2023} also identify (and claim the discovery of) S2 (OCSN 99) and S10 (OSCN 88), which we found in earlier works as Theia 368 by \citet{Kounkel2019} and Volans-Carina by~\citet{Gagne:2018:VolCar}, respectively. Finally, \citetalias{Hunt:2023} has identified fragments of Ratzenboeck~1 (S1; see Fig.\,\ref{fig:cluster_comparison}, top panel) and S12 but does not connect them to a larger structure. Our analysis of the sources' 3D velocity space suggests that these fragments likely correspond to the same stellar structure and should thus not be separated (see Ratzenb\"ock et al. in prep). Disk stream S9 is not recovered by \citetalias{Hunt:2023}. 

The comparison to \citet[hereafter \citetalias{Kounkel2019}]{Kounkel2019} also reveals many overlaps, although we find that many crossmatches appear to describe (to some degree) different stellar aggregates. We find a good alignment between disk streams S2 and Theia 368 (OCSN 99 in \citetalias{Qin:2023}), S3 and Theia 430 (HSC 2303 in \citetalias{Hunt:2023}), and S10 and Theia 424 (Volans-Carina in \citet{Gagne:2018:VolCar}). Comparisons between other disk streams and Theia groups do not yield a precise alignment; for example, Theia 371 (OSCN 87 in \citetalias{Qin:2023} and HSC 2407 in \citetalias{Hunt:2023}) corresponds to the core of S6; however, Theia 371 and S6 strongly disagree on the remaining extent of respective populations. Except for Theia 134 and Theia 508, which we find represent Platais 9 (S4), we find even more extreme mismatches (compared to Theia 371 and S6) where crossmatched groups correspond to a (sometimes vastly) different stellar population or exhibit (high) contamination (e.g., apparent in the color-magnitude diagram), already indicated by other authors\,\citep[see, e.g.,][]{Zucker:2022}. Figure~\ref{fig:cluster_comparison} (middle panel) exemplifies one of these difficult comparisons, in this case, between Theia 435 and Mamajek 2 (S8). Disk streams Ratzenboeck~1 and HSC 2278 (S5) are not recovered by \citetalias{Kounkel2019}. 

We find further crossmatches with the following literature catalogs: \citet{Fuernkranz:2024} recover S2 (Theia 368 in \citetalias{Kounkel2019}) and S6 (OCSN 87 in \citetalias{Qin:2023}), although with substantial contamination (see Fig.\,\ref{fig:cluster_comparison}, bottom panel). \citet{Cantat-Gaudin2020a} and \citetalias{Meingast2021} both contain Platais 9 (our S4) and NGC 2451A (our S7). Compared to the census of \citet{Cantat-Gaudin2020a}, we recover approximately three times the number of sources for Platais 9 and NGC 2451A. The comparison to \citetalias{Meingast2021} yields similar cluster morphologies but an average improvement in the number of identified sources between $30$ and $50\%$. However, the improvement in cluster size is likely due to the improved astrometry of \emph{Gaia} DR3 over DR2 and less stringent error cuts in our work. Finally, the disk stream S10 (OCSN 88 in \citetalias{Qin:2023}) corresponds to the Volans-Carina Association, first discovered by \citet{Gagne:2018:VolCar}, whose source population we have increased approximately tenfold (and roughly doubled against \citetalias{Hunt:2023}). \citet{Moranta:2022} also identify Volans-Carina alongside its corona, where they uncover a total of 141 high-fidelity members (here we find 566 potential members).

% ---------------- Comparison table --------------
\begin{sidewaystable*}
\begin{center}
\begin{small}
\caption{Disk stream comparison against literature catalogs.\vspace{-1.5mm}}
\renewcommand{\arraystretch}{1.4}
\resizebox{0.99\textwidth}{!}{
\begin{tabular}{lllrrrrlrrrrlrrrrlrrrr}

\hline \hline

\multicolumn{17}{c}{Blind search} &
\multicolumn{5}{c}{Targeted search} \\
\cmidrule(lr){1-17}
\cmidrule(lr){18-22}

\multicolumn{2}{c}{This work} &
\multicolumn{5}{c}{\citet{Hunt:2023}} &
\multicolumn{5}{c}{\citet{Kounkel2019}} &
\multicolumn{5}{c}{\citet{Fuernkranz:2024}} &
\multicolumn{5}{c}{\citet{Meingast2021}} \\[-0.8mm]
% \multicolumn{3}{c}{\citet{Cantat-Gaudin2020a}} \\
% \multicolumn{3}{c}{Gagne+2018b} \\

\cmidrule(lr){1-2}
\cmidrule(lr){3-7}
\cmidrule(lr){8-12}
\cmidrule(lr){13-17}
\cmidrule(lr){18-22}
% \cmidrule(lr){15-17}
% \cmidrule(lr){18-20}

% \cmidrule(lr){4-6}
% \cmidrule(lr){7-8}
% \cmidrule(lr){9-10}

\multicolumn{1}{l}{SID} &
\multicolumn{1}{l}{Size} &
\multicolumn{1}{l}{Name} &
\multicolumn{1}{c}{Size\,\tablefootmark{a}} &
\multicolumn{1}{c}{Shared\,\tablefootmark{b}} &
\multicolumn{1}{c}{Gain\,\tablefootmark{c}} &
\multicolumn{1}{c}{$f_{C}$ [\%]\,\tablefootmark{d}} &
\multicolumn{1}{l}{Name} &
\multicolumn{1}{c}{Size} &
\multicolumn{1}{c}{Shared} &
\multicolumn{1}{c}{Gain} &
\multicolumn{1}{c}{$f_{C}$} &
\multicolumn{1}{l}{Name} &
\multicolumn{1}{c}{Size} &
\multicolumn{1}{c}{Shared} &
\multicolumn{1}{c}{Gain} &
\multicolumn{1}{c}{$f_{C}$} &
\multicolumn{1}{l}{Name} &
\multicolumn{1}{c}{Size} &
\multicolumn{1}{c}{Shared} &
\multicolumn{1}{c}{Gain} &
\multicolumn{1}{c}{$f_{C}$} \\

\hline\\[-4mm]
% Submarine
1 & 273 & 
% Hunt (1)
% HSC 2672 & 19.5 & \phantom{0}14 &
\textcolor{teal}{HSC 2672}\,\tablefootmark{f} & \textcolor{teal}{14} & \textcolor{teal}{14} & \textcolor{teal}{19.5} & \textcolor{teal}{100} &
% Kounkel
 - & - & - & - & - &
% Fürnkranz
 - & - & - & - & - &
% Meingast
 - & - & - & - & - \\[0mm]
% Submarine
 &  & 
% Hunt (2)
\textcolor{teal}{Theia 436}\,\tablefootmark{f} & \textcolor{teal}{19} & \textcolor{teal}{19} & \textcolor{teal}{14.4} & \textcolor{teal}{100} & & & & & & & & & & & & & & & \\[2mm]

% ---------------------------
% Submarine
2 & 172 & 
% Hunt
OCSN 99 & 196 & 131 & 1.3 & 73 &
% Kounkel
Theia 368 & 126 & 79 & 2.2 & 79 & 
% Fürnkranz
2 & 128 & 52 & 3.3 & 69 &
% Meingast
 - & - & - & - & - \\[0mm]
% Cantat-Gaudin
 % - & - & - &  \\[-5mm]
% Gagne
 % - & - & - &  \\[-5mm]
% Submarine
 &  & 
% Hunt
 &  &  &  & &
% Kounkel
\textcolor{purple}{Theia 1569}\,\tablefootmark{e} & \textcolor{purple}{124} & \textcolor{purple}{31} & \textcolor{purple}{5.5} & \textcolor{purple}{65} 
& & & & & & & & & & \\[2mm]
% \hline
% ---------------------------
% Submarine
3 & 248 & 
% Hunt (1)
HSC 2303 & 153 & 133 & 1.9 & 93 &
% Kounkel
Theia 430 & 214 & 128 & 1.9 & 74 &
% Fürnkranz
- & - & - & - & - &
% Meingast
 - & - & - & - & - \\[2mm]
% Cantat-Gaudin
 % - & - & - & \\[-1mm]
% Gagne
 % & & & \\[-1mm]
% \hline
% ---------------------------
% Submarine
4 & 454 & 
% Hunt (1)
Platais 9 & 251 & 232 & 2.0 & 96 &
% Kounkel
\textcolor{teal}{Theia 134}\,\tablefootmark{f} & \textcolor{teal}{339} & \textcolor{teal}{233} & \textcolor{teal}{1.9} & \textcolor{teal}{81} &
% Fürnkranz
- & - & - & - & - & 
% Meingast
Platais 9 & 320 & 286 & 1.6 & 93 \\[0mm]
% Cantat-Gaudin
% Platais 9 & 144 & 119 & \\[-5mm]
% Gagne
 % - & - & - &  \\[-5mm]
% Submarine
 &  & 
% Hunt
 &  &  &  &  &
% Kounkel
\textcolor{teal}{Theia 508}\,\tablefootmark{f} & \textcolor{teal}{200} & \textcolor{teal}{77} & \textcolor{teal}{5.9} & \textcolor{teal}{79} &
& & & & & & & & & \\[2mm]
% \hline
% ---------------------------
% Submarine
5 & 203 & 
% Hunt (1)
HSC 2278 & 46 & 40 & 5.1 & 97 &
% Kounkel
- & - & - & - & - & 
% Fürnkranz
- & - & - & - & - & 
% Meingast
- & - & - & - & - \\[2mm]
% Cantat-Gaudin
% - & - & - &  \\[-1mm]
% Gagne
% - & - & - &  \\[-1mm]
% \hline
% ---------------------------
% Submarine
6 & 172 & 
% Hunt
HSC 2407 & 81 & 61 & 2.8 & 90 &
% Kounkel
\textcolor{purple}{Theia 371}\,\tablefootmark{e} & \textcolor{purple}{86} & \textcolor{purple}{37} & \textcolor{purple}{4.6} & \textcolor{purple}{78} & 
% Fürnkranz
\textcolor{purple}{52}\,\tablefootmark{e} & \textcolor{purple}{295} & \textcolor{purple}{70} & \textcolor{purple}{2.5} & \textcolor{purple}{43} & 
% Meingast
- & - & - & - & - \\[2mm]
% Cantat-Gaudin
% - & - & - &  \\[-1mm]
% Gagne
% - & - & - &  \\[-1mm]
% \hline
% ---------------------------
% Submarine
7 & 845 & 
% Hunt
NGC 2451A & 407 & 375 & 2.3 & 90 &
% Kounkel
\textcolor{purple}{Theia 118}\,\tablefootmark{e} & \textcolor{purple}{1857} & \textcolor{purple}{628} & \textcolor{purple}{1.3} & \textcolor{purple}{41} & 
% Fürnkranz
- & - & - & - & - & 
% Meingast
NGC 2451A & 637 & 546 & 1.5 & 90 \\[2mm]
% Cantat-Gaudin
% NGC 2451A & 351 & 312 & \\[-1mm]
% Gagne
% - & - & - &  \\[-1mm]
% \hline
% ---------------------------
% Submarine
8 & 324 & 
% Hunt (1)
Mamajek 2 & 226 & 185 & 1.8 & 89 &
% Kounkel
\textcolor{purple}{Theia 435}\,\tablefootmark{e} & \textcolor{purple}{682} & \textcolor{purple}{152} & \textcolor{purple}{2.1} & \textcolor{purple}{38} &
% Fürnkranz
- & - & - & - & - & 
% Meingast
- & - & - & - & - \\[2mm]
% Cantat-Gaudin
% - & - & - & \\[-1mm]
% Gagne
% - & - & - &  \\[-1mm]
% ---------------------------
% Submarine
9 & 534 & 
% Hunt (1)
- & - & - & - & - &
% Kounkel
\textcolor{purple}{Theia 301}\,\tablefootmark{e} & \textcolor{purple}{1313} & \textcolor{purple}{188} & \textcolor{purple}{2.8} & \textcolor{purple}{32} &
% Fürnkranz
- & - & - & - & - & 
% Meingast
- & - & - & - & - \\[2mm]
% Cantat-Gaudin
% - & - & - & \\[-1mm]
% Gagne
% - & - & - &  \\[-1mm]
% ---------------------------
% Submarine
10 & 566 & 
% Hunt (1)
OCSN 88 & 333 & 303 & 1.9 & 95 &
% Kounkel
Theia 424 & 150 & 90 & 6.3 & 90 & 
% Fürnkranz
- & - & - & - & - & 
% Meingast
- & - & - & - & - \\[2mm]
% Cantat-Gaudin
% - & - & - & \\[-1mm]
% Gagne
% Vol-Car & 0.1 & 54 &  \\[-1mm]
% \hline
% ---------------------------
% Submarine
11 & 196 & 
% Hunt (1)
OCSN 3 & 78 & 71 & 2.8 & 97 &
% Kounkel
\textcolor{purple}{Theia 431}\,\tablefootmark{e} & \textcolor{purple}{297} & \textcolor{purple}{137} & \textcolor{purple}{1.4} & \textcolor{purple}{55} &
% Fürnkranz
- & - & - & - & - & 
% Meingast
- & - & - & - & - \\[2mm]
% Cantat-Gaudin
% - & - & - & \\[-1mm]
% Gagne
% - & - & - &  \\[-1mm]
% \hline
% ---------------------------
% Submarine
12 & 418 & 
% Hunt (1)
\textcolor{teal}{HSC 2327}\,\tablefootmark{f} & \textcolor{teal}{91} & \textcolor{teal}{86} & \textcolor{teal}{4.9} & \textcolor{teal}{99} &
% Kounkel
\textcolor{purple}{Theia 599}\,\tablefootmark{e} & \textcolor{purple}{636} & \textcolor{purple}{258} & \textcolor{purple}{1.6} & \textcolor{purple}{53} & 
% Fürnkranz
- & - & - & - & - & 
% Meingast
- & - & - & - & - \\[0mm]
% Cantat-Gaudin
% - & - & - & \\[-5mm]
% Gagne
% - & - & - &  \\[-5mm]
% Submarine
 &  & 
% Hunt
\textcolor{teal}{HSC 2351}\,\tablefootmark{f} & \textcolor{teal}{34} & \textcolor{teal}{32} & \textcolor{teal}{13.1} & \textcolor{teal}{100} & & & & & & & & & & & & & & & \\[1mm]

\hline
\end{tabular}
} % end resize box
\renewcommand{\arraystretch}{1}
\label{tab:comparison}
\tablefoot{Clusters from literature catalogs are listed in this table if their overlap is at least $10\%$ of the respective smaller cluster. For example, if the disk stream size is $300$ and a given literature cluster has $150$ members, their combined overlap must equal or exceed $15$. We additionally impose a minimum overlap size of $10$ to exclude chance alignments between unrelated clusters.
\tablefoottext{a}{The ``size'' reports the absolute number of sources in respective literature searches.}
\tablefoottext{b}{The ``shared'' column reports the size of the intersection/overlap between sources from our selection and the corresponding literature catalog cluster.}
\tablefoottext{c}{The ``gain'' column shows how much our survey increases source counts compared to a specific literature catalog. It compares our group size to the intersection size, representing confirmed samples in the literature. The gain is always greater than one and adjusts for potential contamination in the literature sample (see Fig.~\ref{fig:cluster_comparison}). This statistic becomes meaningful under the assumption of low contamination in our sample.}
\tablefoottext{d}{The column ``$f_C$'' reports an estimation of the completeness fraction (in \%) of our sample based on a given reference sample from the literature. We compute the completeness $f_C$ by first considering the incompleteness fraction (1 - $f_C$). We define incompleteness as the number of literature sources we are unable to find divided by the union of our sample and the literature sample. The union provides an estimate of the total number of potential cluster members relative to a given literature selection. Note, however, that the incompleteness fraction is only meaningful as a comparison metric as it depends heavily on the quality of the literature selection and does not take into account the reference's contamination and completeness.}
\tablefoottext{e}{(\textcolor{purple}{contaminated}) The corresponding literature cluster partially overlaps with our selection but appears to be (sometimes significantly) contaminated or describes a (sometimes vastly) different stellar population. See examples in Fig.\,\ref{fig:cluster_comparison}.}
\tablefoottext{f}{(\textcolor{teal}{merge proposed}) Several literature clusters suggest that our selection of two disk streams (S1, S4, and S12) should actually be split into multiple clusters. Analyzing the 3D velocity and HRD distribution of these stream fragments (see Ratzenb\"ock et al. in prep), we conclude that these fragments likely correspond to the same stream and should thus not be separated.}
}
\end{small}
\end{center}
\end{sidewaystable*}
% -------------------------------------------------------

\section{Velocity dispersion estimation}\label{app:vel_disp}
In this section we provide some additional details on the velocity dispersion estimation alongside a discussion on the size of the derived dispersion estimates. 

\subsection{Modified extreme deconvolution}
We employed the XD method from \citet{Bovy:2011} to approximate the noise-free distribution of sources in the Galactic Cartesian velocity space (\emph{UVW}). Using XD, we aim to minimize the impact of large radial velocity measurement errors on the resulting 3D velocity dispersion.

The term ``extreme'' in XD highlights its ability to reconstruct the underlying density function even when each source has a unique Gaussian noise covariance matrix. XD utilizes an expectation-maximization (EM) algorithm \citep{Dempster1977} that iteratively maximizes the likelihood of a GMM representing the noise-free distribution, convolved with the individual error covariances of all measurements.

We aim to use XD to infer a deconvolved density in the 3D Galactic Cartesian velocity space for each of the 12 disk streams. To achieve this, we transform the observations (proper motions and radial velocities) and their corresponding error covariances into \emph{UVW} space. This involves computing the Jacobian $J$ of the transformation between on-sky velocities ($\mu_{\alpha}$, $\mu_{\delta}$, $v_r$) and space velocities $(U, V, W)$. The Jacobian was then used to transform the observed covariance matrix, $C_{\text{ICRS}}$, into the error covariance matrix in Galactic Cartesian velocity space, $C_{\text{Gal}}$, as follows\footnote{Both the covariances $C_{\text{ICRS}}$ and $C_{\text{Gal}}$ and the Jacobian depend on a source's on sky position (ra, dec) and its parallax. Thus, we operate Eq.(\ref{eq:trafo}) for each source in the catalog.}:

\begin{equation}\label{eq:trafo}
    C_{\text{Gal}} = J\,C_{\text{ICRS}}\,J^{T}.
\end{equation}

Assuming approximately Gaussian-distributed 3D velocities, we represent the signal as a single Gaussian distribution. Given the expected nonzero contamination, we explicitly model the contaminating sources. Contamination in this context refers to sources whose space motions and corresponding uncertainties are incompatible with the signal distribution. Adding a second Gaussian component to the mixture distribution effectively describes the background contamination. To prevent one Gaussian component from collapsing, we ensure that the background component accounts for at least $5\%$ of the observations, discouraging a single mixture component from modeling the entire distribution. 

We initialized the mean and covariance matrix of the signal component with values computed from the progenitor stream. The parameters of the background component were initialized as the mean velocity and empirical covariance estimate of the entire local 6D dataset (see Sect.\,\ref{sect:data}), which serves as the basis for each stream search. To ensure that the background component only models uncorrelated and long-range structures, we constrained it to have a diagonal covariance matrix with each diagonal element exceeding a velocity dispersion of 10\,km\,s$^{-1}$. This constraint prevents the background component from ``collapsing'' and modeling parts of the signal distribution.

\begin{figure*}
    \centering
    \includegraphics[width=0.94\textwidth]{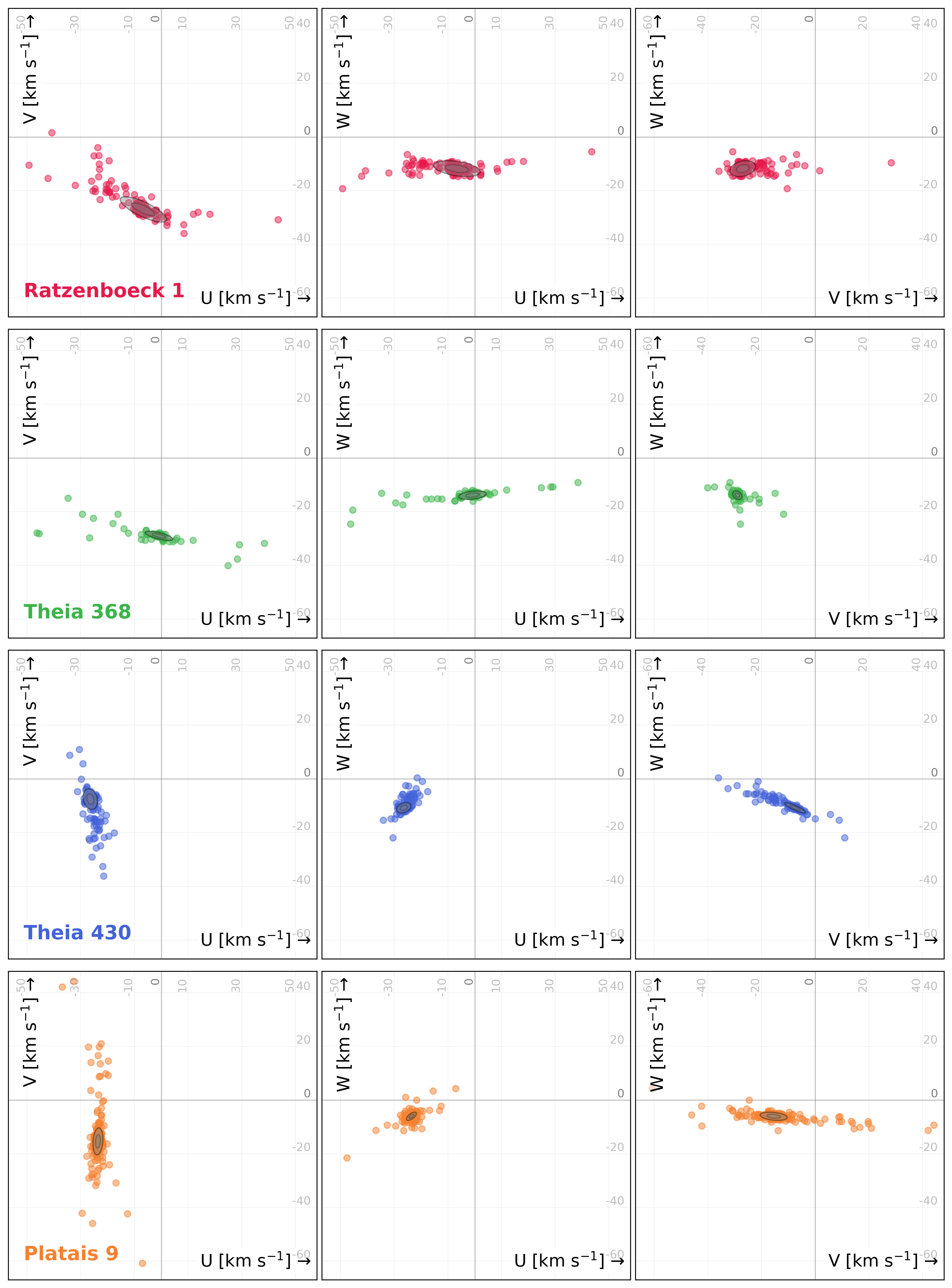}
    \caption{\emph{Gaia} Cartesian velocity distributions (\emph{UVW}) of the 12 recovered disk streams. The black ellipses show the 1$\sigma$ and 2$\sigma$ confidence regions inferred by the XD procedure (see Appendix\,\ref{app:vel_disp} for more details). The XD procedure can effectively ignore the large line-of-sight velocity errors, which produce the pronounced elongation feature seen in most scatter plots. This figure shows the \emph{UVW} distribution of disk streams S1 - S4.}
    \label{fig:uvw_1}
\end{figure*}

\begin{figure*}
    \centering
    \includegraphics[width=0.95\textwidth]{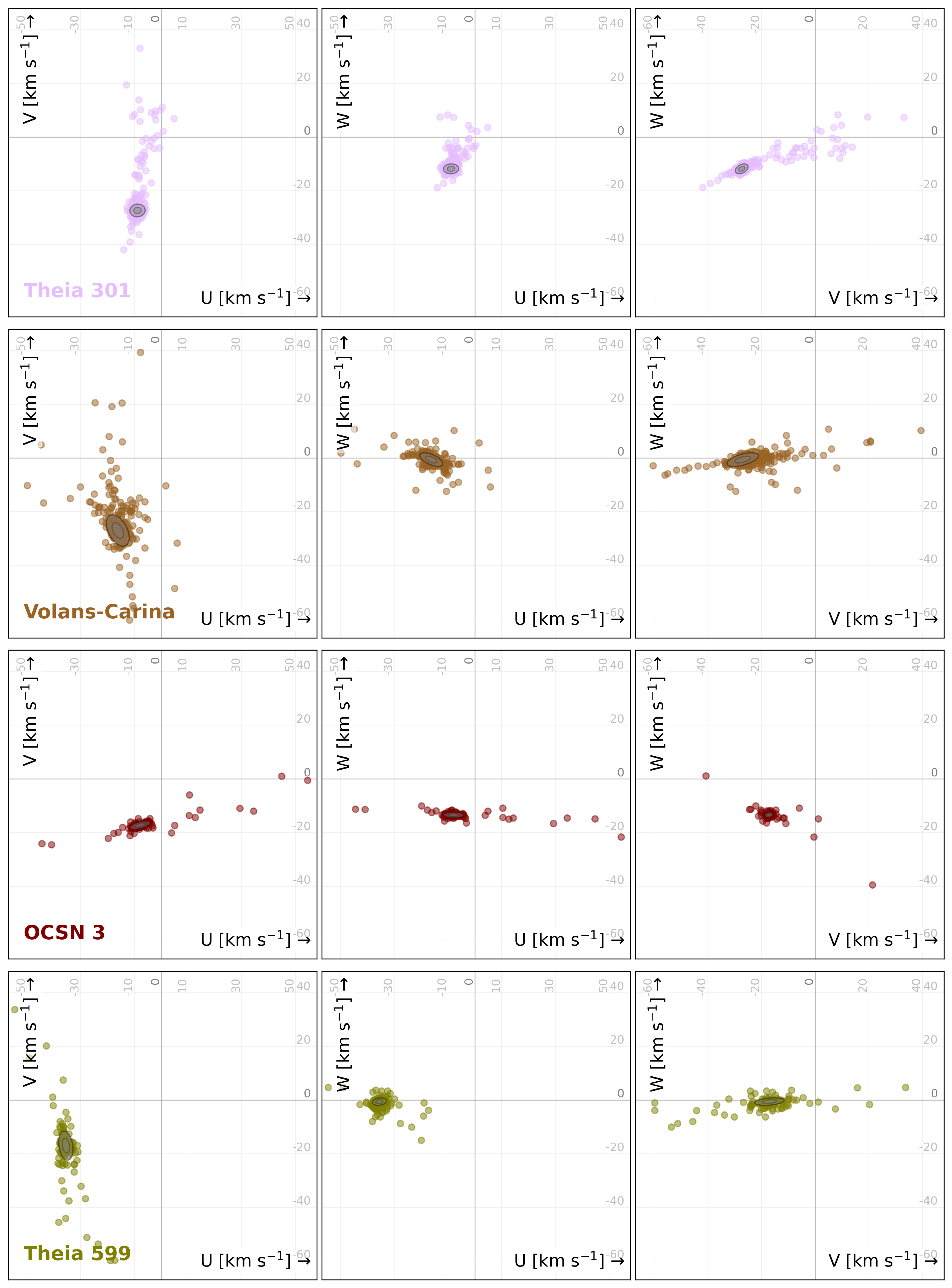}
    \caption{Same as Fig.\,\ref{fig:uvw_1}, but for disk streams S9 - S12.}
    \label{fig:uvw_3}
\end{figure*}

Using these constraints, we empirically find that the signal component in the GMM effectively captures the space velocity distribution of each disk stream. Figures\,\ref{fig:uvw_1}\,-\,\ref{fig:uvw_3} show the distribution of sources in Cartesian velocity space alongside 1- and 2-$\sigma$ covariance ellipses of the normal signal component inferred by the XD procedure. Table\,\ref{tab:group_stats} presents the corresponding mean space motion and the 3D velocity deviation, $\sigma_{3D}$. The 3D velocity deviation is defined as $\sigma_{3D}^2 = \sigma_U^2 + \sigma_V^2 + \sigma_W^2$, which corresponds to the trace of the covariance matrix.

Finally, we derived an estimate of contamination using this procedure. Our GMM explicitly models the density as a mixture of signal and background components, allowing us to obtain the contamination estimate directly from the mixture weight assigned to the background component. This estimate inevitably includes biases introduced by the discussed model choices, notably that the contamination estimate cannot fall below $5\%$. Nevertheless, we consider this estimate a reasonable first-order approximation of the contamination fraction within our sample. We find the contamination ranges from $5\%$ to $17\%$, with a mean contamination of $9\%$.

\subsection{Discussion}
Our XD analysis reveals that the disk stream populations are dynamically cold, with 3D velocity dispersions between $2.1$ and $5.1$\,km\,s$^{-1}$. Velocity dispersions below $\sim 5$\,km\,s$^{-1}$ are notably low for structures spanning several hundred parsecs. Especially since, in Galactic heliocentric velocities, the contribution from Galactic rotation for these large structures is not negligible, amounting between 1 and 3\,km\,s$^{-1}$. 

Changing to Galactocentric cylindrical or action-angle coordinates can provide a more natural way to estimate the remaining intrinsic dispersion as it removes the contribution of Galactic rotation. Our tests conclude that the clustering pipeline is robust to the change of velocity coordinate system (especially for velocities in Galactocentric and Galactic coordinates), with results finding no significant systematic differences in the resulting streams’ morphology (length, location, velocity dispersion). Ultimately, we opted to use and report our findings in the heliocentric Galactic Cartesian coordinate system primarily because it facilitates direct comparisons with other cluster studies that mainly use the same coordinate system.

Comparison to literature values, for example, the fiducial disk stream Meingast-1 (Pisces Eridanus) with a length of approximately 400\,pc, has a 3D velocity dispersion in Galactic heliocentric velocities of $\sim 3$\,km\,s$^{-1}$~\citep{Meingast2019}. We also empirically find velocity dispersions below 5\,km\,s$^{-1}$ (in Galactic heliocentric velocities) when analyzing other elongated stellar structures with extents of 100 – 200 pc like cluster coronae~\citep{Meingast2021, Moranta:2022} and tidal tails~\citep{MeingastAlves2019, Roeser2019, Tarricq:2022}. As discussed in Sect.~\ref{sect:dissolution}, we believe the identified disk stream sample has a similar origin to coronae and tidal tails -- namely Galactic tidal forces and differential rotation -- which might be responsible for such a small velocity dispersion. In future work, when a more comprehensive sample is available, we aim to investigate the kinematic profile of disk streams and in more detail.

Another major factor contributing to small velocity dispersion measurements (also across similar literature examples) is selection bias. The large size of identified structures means that the recovered members are embedded in a large background population. Members from the tail of the velocity distribution that are a few km\,s$^{-1}$ away from the central overdensity (in whichever velocity coordinate system) will not have enough contrast over the dominating background (i.e., a S/N of $\sim 1$) to be detected by density-based clustering methods that are typically employed to search for these structures.

Finally, the XD algorithm may bias the results by favoring a more ``compact'' signal component over a broader or more dispersed one, potentially leading to underestimating the true velocity dispersion. Since we employed a two-component mixture model with restrictions placed on one component, the model may be prone to overfitting, potentially leading to an underestimation of the signal component’s size. Evidence supporting this hypothesis includes contamination estimates from the XD procedure, which are, on average, approximately $2\%$ higher than those derived from the 3D velocity histogram method (see Appendix~\ref{app:contamination}). 

To address the potential underestimation of the velocity dispersion, we provided a reference estimate using a robust measure, the MAD, to calculate the 3D velocity dispersion. The estimation was done including only sources with radial velocity errors below 1\,km\,s$^{-1}$ to minimize biases from large measurement uncertainties. This approach increased the average 3D velocity dispersion from 2.9\,km\,s$^{-1}$ to 4.6\,km\,s$^{-1}$, as shown in Table~\ref{tab:group_stats}.

\subsection{Kinematic independence of disk streams}\label{app:kinematic_pop_analysis}
This study's derivation of disk stream densities assumes that the 12 identified streams are independent structures. However, the spatial overlap (see Fig.~\ref{fig:3d_view}) between several of these streams raises the question of whether some might represent fragments of larger structures instead. Concretely, we find three groups of disk streams that are, at least partially, co-spatial: (1) Theia 430 and HSC 2278, (2) the four disk streams Theia 371/OCSN 87, NGC 2451A, Theia 301, and OCSN 3, and (3) Ratzenboeck~1, Theia 368, and Mamajek 2. 

To investigate this, we examined the kinematic and spatial properties of the identified disk streams. Pairwise comparisons between disk streams reveal that most streams exhibit significantly distinct kinematics, except for one notable case (see below). Using the estimated mean and covariance matrix of each stream’s 3D velocity distribution obtained through the XD procedure, we computed pairwise Mahalanobis distances between the streams’ 3D velocities. The Mahalanobis distance quantifies the separation between two distributions in units of standard deviations, accounting for the shape of their covariance ellipsoids. Distances above 3 typically indicate statistically significant differences in kinematics. As the Mahalanobis distance depends on the covariance matrix of each stream, it is inherently asymmetric.

\begin{figure}
    \centering
    \includegraphics[width=0.98\columnwidth]{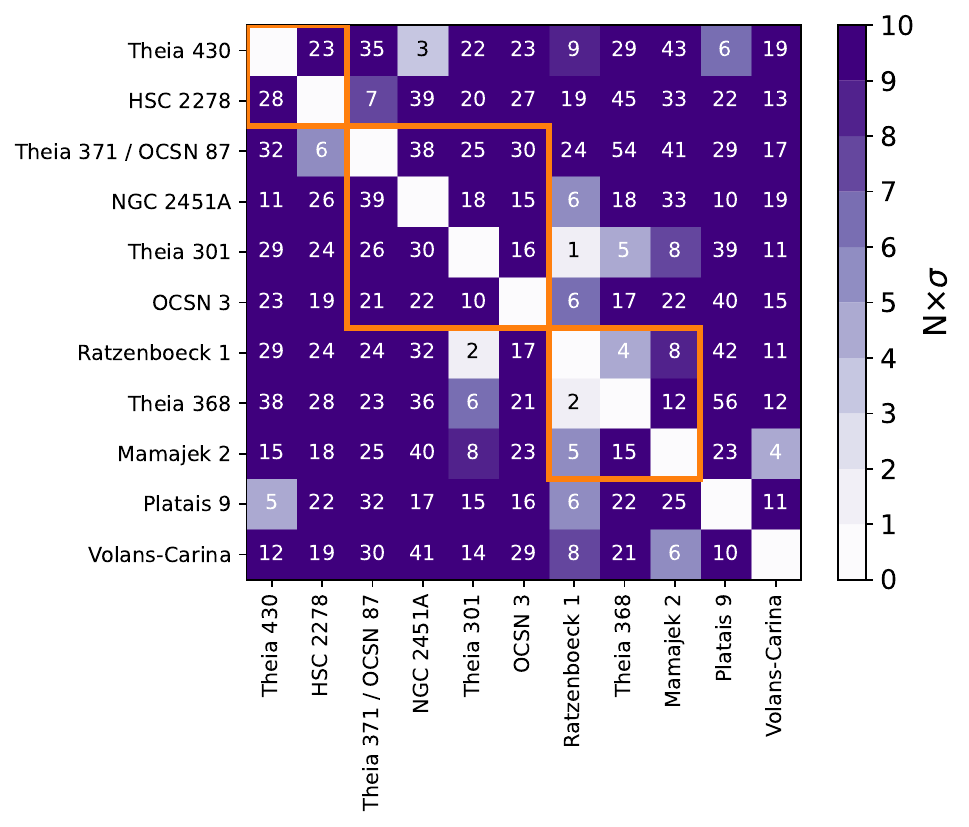}
    \caption{Pairwise Mahalanobis distances between the 3D velocities of all identified disk streams. The Mahalanobis distance quantifies the separation between two streams in units of standard deviations, accounting for the covariance structure of their velocity distributions. Distances greater than 3 (highlighted in darker shades) indicate statistically significant kinematic differences. Orange squares denote stream pairs that partially overlap in 3D spatial volume. The disk streams Ratzenboeck~1 and Theia~386, despite their spatial and kinematic proximity, show evidence of being distinct structures based on their 3D morphology and orbital histories, as discussed in the text.}
    \label{fig:kinematic_differences}
\end{figure}

The pairwise Mahalanobis distances, rounded for clarity, are shown in Fig.~\ref{fig:kinematic_differences}. The rows and columns of the pairwise distance matrix are ordered such that disk streams that are spatially close are next to each other for ease of comparison. Orange squares in the figure highlight streams that (at least partially) overlap in 3D spatial volume. This analysis reveals significant kinematic differences among all but one pairwise comparison, pointing toward co-spatial but inherently independent coeval structures. This claim is substantiated by large age differences among individual clusters in these three groups. 

We find one case that needs further investigation, which is Ratzenboeck~1 and Theia 386. These two disk streams are kinematically similar, hinting at fragments of a larger structure. However, closer examination of their 3D morphology and velocity properties provides a more nuanced picture. The two streams show distinct overdensities in \emph{XYZ} space, and their absolute velocity values point toward separate entities. Specifically, Theia 386, located further along in Galactic rotation, exhibits a relative velocity that points ``backward'' toward Ratzenboeck~1. This velocity signature indicates a contractive motion, which goes against the expected effects of Galactic tidal forces and differential rotation. To substantiate this claim, we performed a traceback analysis, integrating their orbits 20\,Myr into the past and future\footnote{We used the \texttt{galpy} Python library \citep{galpy_15} with the default parameters. This parametrization uses the \texttt{MWPotential2014} as the Galactic potential, the solar motion relative to the local standard of rest from \citet{Schoenrich_10_MNRAS}, ($UVW_{\mathrm{\odot}, LSR}$) = (-11.1, 12.24, 7.25)\,km\,s$^{-1}$, and the solar position relative to the Galactic center of ($XYZ_{\mathrm{G}}$)~=~(8122.0, 0.0, 20.8)\,pc \citep{GravityColl_18_AA, Bennett_Bovy_19_MNRAS}.}. This analysis reveals that both streams diverge when their orbits are integrated into the past, providing further evidence that they are two separate, coeval populations.

Despite these differences, the streams share similar, though not identical, 3D velocities and ages. This suggests a potential joint formation scenario analogous to the substructure observed in OB associations such as Sco-Cen or Orion. Their similarities may reflect a common origin or a shared dynamical history within a larger parent structure.

\section{Contamination estimates}\label{app:contamination}
Identifying stellar streams and clusters in a local volume of the Milky Way is inherently challenging due to contamination from the dominant background of field stars in the same phase-space volume. This issue is particularly pronounced for low-density structures encompassing large spatial regions, as even with advanced clustering algorithms, false positives can arise when background stars exhibit positions and velocities that coincidentally match those of stream members. 

We addressed these concerns by estimating contamination rates through an approach independent of the clustering and XD procedures. Specifically, we estimated the contamination fraction by comparing the identified stream members with co-spatial \emph{Gaia} sources serving as a background population from the same volume. In contrast to the contamination estimate from the XD procedure, which constitutes rather an outlier approximation (see Appendix~\ref{app:vel_disp}), we explicitly considered the background distribution of field stars and their phase-space density.  

Identifying the background population for each stream is a task of finding all co-spatial \emph{Gaia} sources that share the same \emph{XYZ} volume as respective stream members. These sources serve as a reference sample, which should be distributed differently in phase space and the observational HRD if the stream members are genuine. 

We used OCSVM to estimate the support of the positional distribution of each disk stream (i.e., its extent) in Galactic heliocentric coordinates using the identified members. These contours, shown in Fig.~\ref{fig:3d_view}, were used to define the 3D spatial boundaries of each stream. The background population for each stream was subsequently identified as \emph{Gaia} sources from the 500\,pc search data (see Sect.~\ref{sect:data}) that lie within these 3D volumes but exclude the respective stream members. The median background population size of each stream is approximately 20\,000 sources, with a minimum background size of 3\,000 and a maximum background size of 100\,000 sources. This sample of background stars allows us to quantify contamination and S/N levels by comparison with the identified stream candidate members. The subsequent sections provide detailed results from these analyses and discuss their implications for the reliability of the identified disk streams.

\subsection{CMD test}\label{app:cmd_test}
Here, we aim to assess whether the stream populations (see Fig.~\ref{fig:hrd}) are significantly narrower in the color-magnitude diagram (CMD) compared to a random sample from the respective background population. We tested the hypothesis that the distribution of sources in the CMD closely follows a single isochronal curve. If the recovered stream members show a significantly narrower pattern around an isochrone curve than the background population, the contamination likely does not dominate the selected sample.

To test whether these two distributions differ drastically in the CMD, we drew ten bootstrap samples (with replacement) from the disk stream members and the corresponding background populations. The bootstrap sample size is chosen to represent the size of respective streams. For each sample, we computed the sources' closest (and signed) distances to an isochronal curve. For the disk streams, these distances were computed relative to the best fitting isochrone determined in Sect.~\ref{sect:results} (see Table~\ref{tab:group_stats} and Fig.~\ref{fig:hrd}). For the background populations, a new best-fitting isochrone was computed for each bootstrap sample to avoid biases in the comparison. To mitigate the biasing effects of outliers, we removed likely binaries via a cut in the RUWE parameter and white dwarf candidates via a cut in the CMD following \citet{Golovin:2014}:

\begin{equation}\label{eq:cmd_sel}
\begin{split}
    M_G &> 10 + 2.5 \times (G_{BP} - G_{RP})\\
    G_{BP} - G_{RP} &< 1.9\\
    \text{RUWE} &> 1.4
\end{split}
,\end{equation}

Each bootstrap draw resulted in two distributions of (signed) distances: one from the stream members and one from the background population. Across all draws, the signed distance distributions of disk streams have smaller variances than respective background samples. To determine whether the stream variances are significantly smaller and, thus, its members are more compactly distributed around the best fitting isochronal curve, we employed Levene’s test~\citep{Levene:1960}. Levene’s test evaluates the equality of variances between two distributions. It was chosen due to its robustness against deviations from normality. If the p-value from Levene’s test falls below a significance level (e.g., 5\%), indicating that the differences in variance are unlikely a result of random sampling from populations with equal variances, thereby supporting our hypothesis.

From the ten bootstrap samples, we obtained ten p-values for each comparison. Using these ten tests, we performed a combined test of the null hypothesis that no p-value is significant. We applied the harmonic mean p-value method \citep{Wilson:2019}, which is more robust to dependences among p-values than, for example, the \citet{fisher:1934} method, enhancing its reliability in this context.

Our analysis shows that for all but one disk stream (Theia 599), the hypothesis of equal variances can be rejected at a $2 \sigma$ level. For most streams, this hypothesis is rejected at a $3 \sigma$ level, and for some, even at a $5 \sigma$ level, strongly supporting the conclusion that the stream members are more tightly clustered around the isochrone than their respective background populations.

\subsection{3D velocity test}
To explicitly quantify the contamination of each identified disk stream, we performed a velocity-based analysis leveraging the significant differences in velocity dispersion between stream members and the background population. While approximately 10\,000 -- 100\,000 background sources share the same volume alongside selected stream members, the background exhibits a much larger velocity dispersion than the stream members themselves, characterized by velocity dispersions of approximately 5 km\,s$^{-1}$ or less. Hence, the stream's probability density function is densely concentrated in a small region of the \emph{UVW} space, while the background population's probability density function is expected to be relatively diffuse. In the following, we expand on this idea to quantify each stream's S/N and contamination rate. 

\begin{figure}
    \centering
    \includegraphics[width=0.8\columnwidth]{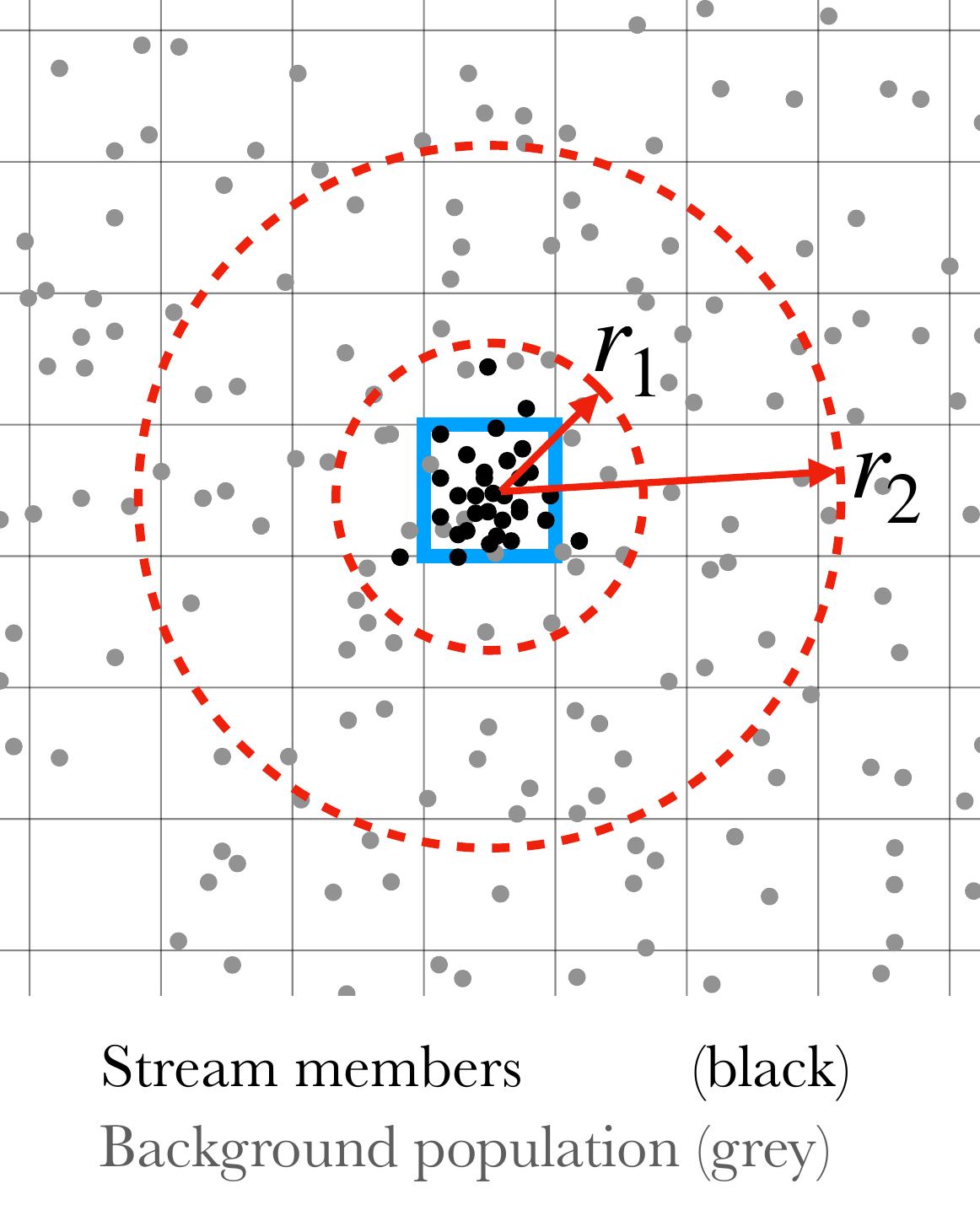}
    \caption{Schematic representation of the procedure used to estimate contamination in velocity space. The figure highlights the central voxel (blue) centered on the bulk velocity of stream members, where their density (black dots) is maximal. Background sources (gray points) are distributed more broadly across velocity space. This spatial separation in \emph{UVW} velocity space enables the estimation of the expected background density at the stream signal location, which is used to calculate each stream's S/N. The red concentric circles illustrate the neighborhoods across which the expected background number density is estimated.}
    \label{fig:contamination_est}
\end{figure}

Using the Galactic heliocentric velocity space components \emph{U}, \emph{V}, and \emph{W} within the range of $(-100, 100)$ km\,s$^{-1}$, we divided this space into bins to construct a 3D histogram. We chose a bin size of 5 km\,s$^{-1}$ along each axis, which guarantees that most stream members fall into one or very few central voxels due to a comparable velocity dispersion for each stream. In contrast, the background population, with its likely more extensive velocity spread, covers a significantly larger volume. For a total of 64\,000 voxels in the defined \emph{UVW} range, even the largest background population ($N \sim 100\,000$) has an average density of only a few stars per voxel, assuming a uniform distribution. Although the background is certainly not uniformly distributed, its broader dispersion assumes a likely lower probability mass in the specific regions (i.e., voxels) occupied by the stream members. Figure~\ref{fig:contamination_est} shows a schematic overview of this procedure, where stream members (in black) are tightly distributed and, thus, predominantly lie in a small volume in velocity space (the blue pixel). In contrast, background sources (gray points) cover the space more uniformly. Therefore, the expected number density of background sources at the signal location is likely significantly lower than the signal density. 

We estimated the expected background density at the stream's bulk (i.e., mean) motion location to quantify the separation between the stream members and the background. Combining this with the signal density, we calculated each stream's S/N. Using the background sources, the expected background at the mean stream velocity was determined via the following procedure. First, we determined the number density in the ``central voxel'' centered on the stream's bulk velocity, denoted as the signal number density. Figure~\ref{fig:contamination_est} shows the central voxel highlighted in blue in which the number density of stream members (black points) is maximal. Second, the expected number density of background sources at the signal location is determined by averaging the voxel count across neighboring cells. Since the background source count varies slightly from voxel to voxel, we averaged the bin counts across multiple cells to obtain a more robust estimate. Specifically, we selected voxels around the central voxel in a growing concentric sphere and computed an expected source count for all voxels whose center location is inside this sphere. We schematically depict this procedure in Fig.~\ref{fig:contamination_est} via the red circles. We selected a minimum radius of 5 km\,s$^{-1}$ that includes the central voxel and its immediate neighbors and a maximum radius of 30 km\,s$^{-1}$. This results in six unique estimates of the expected number density of background sources at the signal location. We obtain an S/N estimate by dividing the signal number density by the expected background number density for each radius value. 

\begin{figure*}
    \centering
    \includegraphics[width=0.95\textwidth]{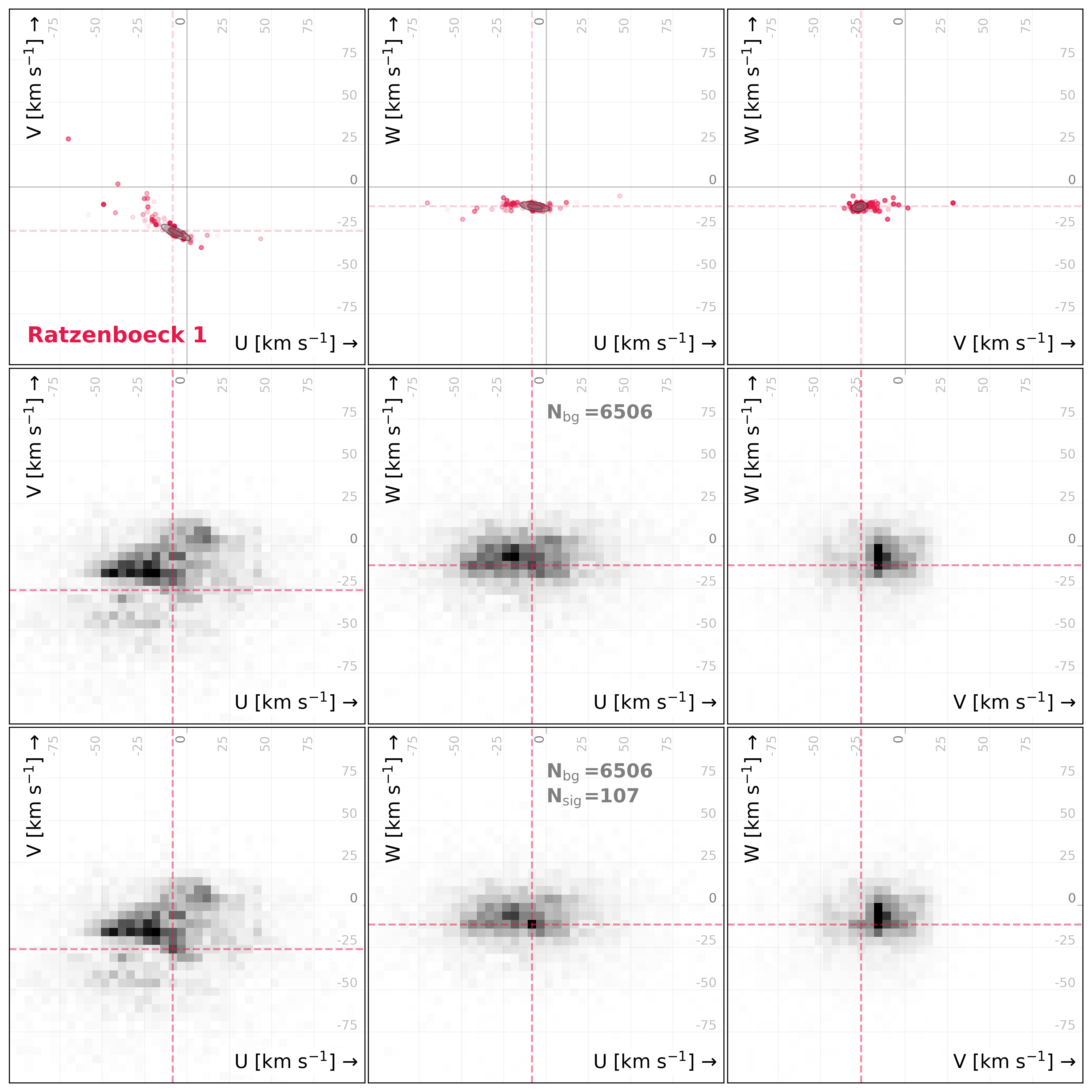}
    \caption{S/N estimation procedure applied to the Ratzenboeck~1 stream. The 3D histogram in the middle and bottom row is shown as three marginalized 2D histograms, showing \emph{U-V}, \emph{U-W}, and \emph{V-W} combinations of the disk stream Ratzenboeck~1 and its respective background population. The top row displays the individual stream members in red (akin to Fig.~\ref{fig:uvw_1}). The middle and bottom rows show the distribution of background sources and the combined sample of signal and background, respectively. The color map represents the number density. Dark regions symbolize high and light gray regions low number densities. Each histogram is normalized and integrates to unity. The horizontal and vertical dashed line indicates the bulk (i.e., the median) velocity of the disk stream. The voxel, which contains most of the disk stream, has a significant number density increase when adding the identified stream candidate members, as shown in the bottom row. The 3D histogram plots showing the other 11 disk streams are provided online via Zenodo, using the following \href{https://zenodo.org/records/14278685}{link}.
}
    \label{fig:uvw_contamination}
\end{figure*}

Figure~\ref{fig:uvw_contamination} displays the 3D histogram as three marginalized 2D histograms, showing \emph{U-V}, \emph{U-W}, and \emph{V-W} combinations of the disk stream Ratzenboeck~1 and its respective background population. The top row shows only the stream members in red, while the middle and bottom rows show the distribution of background samples and the combined sample of signal and background, respectively. The color map represents the number density where dark gray regions symbolize high and light gray regions low number densities. Each histogram is normalized such that the total area integrates to unity. The horizontal and vertical dashed red lines indicate the bulk (i.e., the mean) velocity of the disk stream. The voxel, which contains most of the disk stream members, has a significant number density increase when adding the identified stream candidate members, as shown in the bottom row. This results in an approximate S/N of $10 \pm 2$ for Ratzenboeck~1. The remaining 3D histogram plots are provided online via Zenodo, using the following \href{https://zenodo.org/records/14278685}{link}.

This analysis revealed an average S/N of 27 across all disk streams, with individual values ranging from 5 to 114. These S/N values translate into average contamination estimates of $7 \pm 4\%$. These estimates align well with contamination fractions inferred by XD (see Appendix~\ref{app:vel_disp}). Table~\ref{tab:group_stats} provides each stream's mean S/N estimates alongside its standard deviation from different radius values. A mean S/N of 27 highlights the robustness of the stream selection process, further supporting the result from our CMD test (see Appendix~\ref{app:cmd_test}) that contamination does not dominate the identified disk stream catalog.

\begin{figure*}
    \centering
    \includegraphics[width=0.95\textwidth]{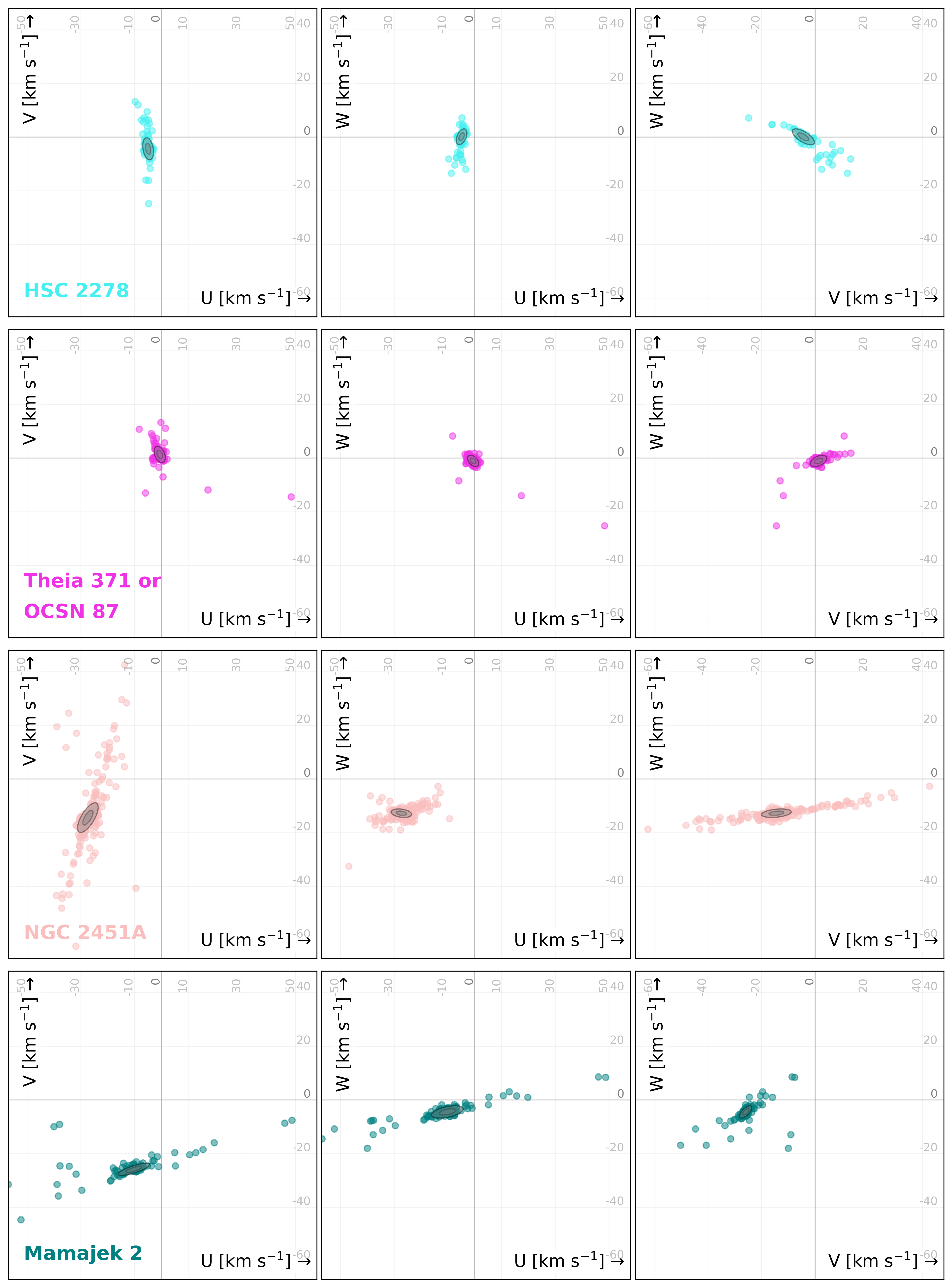}
    \caption{Same as Fig.\,\ref{fig:uvw_1}, but for disk streams S5 - S8.}
    \label{fig:uvw_2}
\end{figure*}

\section{Auxiliary tables and figures}\label{app:aux_tables}

In Table~\ref{tab:member-catalog} we give an overview of the contents of the source-level catalog containing all identified disk stream members as selected in this work alongside membership labels. The full version of the table is available online as a machine-readable version. The phase space coordinates ($XYZUVW$) are defined within the heliocentric Galactic Cartesian coordinate system where the $X$-axis grows positive toward the Galactic center, the $Y$-axis grows positive in direction of Galactic rotation, and the $Z$-axis grows positive toward the Galactic north pole.

\begin{table}[!h] 
\begin{center}
\begin{small}
\caption{Catalog of the identified disk stream members, labeled by membership.}
\renewcommand{\arraystretch}{1.2}
\resizebox{0.85\columnwidth}{!}{%
\begin{tabular}{ll}
\hline \hline 
\multicolumn{1}{l}{Column name} &
\multicolumn{1}{l}{Column description} \\
\hline 
\texttt{source\_id} & \textit{Gaia} DR3 source identification number  \\
\texttt{SID} & Membership label; see Table~\ref{tab:group_stats}   \\
\texttt{name} & Literature name; see Table~\ref{tab:group_stats} \\
\texttt{ra} & RA (deg)  \\
\texttt{dec} & Declination (deg)  \\
\texttt{X} & $X$ position (pc) \\
\texttt{Y} & $Y$ position (pc) \\
\texttt{Z} & $Z$ position (pc) \\
\texttt{U} & $U$ velocity component (km\,s$^{-1}$) \\
\texttt{V} & $V$ velocity component (km\,s$^{-1}$) \\
\texttt{W} & $W$ velocity component (km\,s$^{-1}$) \\
\hline
\end{tabular}
} % end resize box
\renewcommand{\arraystretch}{1}
\label{tab:member-catalog}
\tablefoot{The full machine-readable version of the catalog is available online, while a column overview is given here. We list all relevant derived parameters. Original \textit{Gaia} DR3 parameters can be queried from the \textit{Gaia} Archive by using the \texttt{source\_id}.
}
\end{small}
\end{center}
\end{table}

\end{appendix}

\end{document}